\documentclass[12pt,a4paper]{article}

\usepackage{ifthen} 
\newboolean{pdflatex}
\setboolean{pdflatex}{true} 

\newboolean{articletitles}
\setboolean{articletitles}{true} 

\newboolean{uprightparticles}
\setboolean{uprightparticles}{false} 

\newboolean{inbibliography}
\setboolean{inbibliography}{false} 

\def\papertitle{More visualisation of decay-time-dependent asymmetries in multibody $B$-meson decays} 

\usepackage{microtype}
\usepackage{lineno}  
\usepackage{xspace} 
\usepackage{caption} 

\usepackage{graphicx}  
\usepackage{color}
\usepackage{colortbl}
\graphicspath{{./figs/}} 

\usepackage{amsmath} 
\usepackage{amssymb}
\usepackage{amsfonts}
\usepackage{upgreek} 

\newcommand*\patchAmsMathEnvironmentForLineno[1]{%
\expandafter\let\csname old#1\expandafter\endcsname\csname #1\endcsname
\expandafter\let\csname oldend#1\expandafter\endcsname\csname
end#1\endcsname
 \renewenvironment{#1}%
   {\linenomath\csname old#1\endcsname}%
   {\csname oldend#1\endcsname\endlinenomath}%
}
\newcommand*\patchBothAmsMathEnvironmentsForLineno[1]{%
  \patchAmsMathEnvironmentForLineno{#1}%
  \patchAmsMathEnvironmentForLineno{#1*}%
}
\AtBeginDocument{%
\patchBothAmsMathEnvironmentsForLineno{equation}%
\patchBothAmsMathEnvironmentsForLineno{align}%
\patchBothAmsMathEnvironmentsForLineno{flalign}%
\patchBothAmsMathEnvironmentsForLineno{alignat}%
\patchBothAmsMathEnvironmentsForLineno{gather}%
\patchBothAmsMathEnvironmentsForLineno{multline}%
\patchBothAmsMathEnvironmentsForLineno{eqnarray}%
}

\usepackage{hyperref}    
\usepackage[all]{hypcap} 

\usepackage{xspace} 
\usepackage{upgreek}

\def\MagUp {\mbox{\em Mag\kern -0.05em Up}\xspace}

\ifthenelse{\boolean{uprightparticles}}%
{

 \def\Pmu         {\ensuremath{\upmu}\xspace}

 \def\Ppi         {\ensuremath{\uppi}\xspace}

 \def\Ppsi        {\ensuremath{\uppsi}\xspace}

 \def\PDelta      {\ensuremath{\Delta}\xspace}                 
 \def\PXi         {\ensuremath{\Xi}\xspace}                 
 \def\PLambda     {\ensuremath{\Lambda}\xspace}                 
 \def\PSigma      {\ensuremath{\Sigma}\xspace}                 
 \def\POmega      {\ensuremath{\Omega}\xspace}                 
 \def\PUpsilon    {\ensuremath{\Upsilon}\xspace}
 \let\oldPi\Pi
 \def\PPi         {\ensuremath{\oldPi}\xspace}

 \def\PB      {\ensuremath{\mathrm{B}}\xspace}                 
                  
 \def\PD      {\ensuremath{\mathrm{D}}\xspace}

 \def\PJ      {\ensuremath{\mathrm{J}}\xspace}                 
 \def\PK      {\ensuremath{\mathrm{K}}\xspace}

 \def\Pi      {\ensuremath{\mathrm{i}}\xspace}

 \def\Ps      {\ensuremath{\mathrm{s}}\xspace}

 \def\thebaroffset{0.0em}
}
{

 \def\Pmu         {\ensuremath{\mu}\xspace}

 \def\Ppi         {\ensuremath{\pi}\xspace}

 \def\Ppsi        {\ensuremath{\psi}\xspace}                 
                  
 \mathchardef\PDelta="7101
 \mathchardef\PXi="7104
 \mathchardef\PLambda="7103
 \mathchardef\PSigma="7106
 \mathchardef\POmega="710A
 \mathchardef\PUpsilon="7107
 \mathchardef\PPi="7105
                  
 \def\PB      {\ensuremath{B}\xspace}                 
                  
 \def\PD      {\ensuremath{D}\xspace}

 \def\PJ      {\ensuremath{J}\xspace}                 
 \def\PK      {\ensuremath{K}\xspace}

 \def\Pi      {\ensuremath{i}\xspace}

 \def\Ps      {\ensuremath{s}\xspace}

 \def\thebaroffset{0.18em}
}
\newcommand{\offsetoverline}[2][\thebaroffset]{\kern #1\overline{\kern -#1 #2}}%

\makeatletter
\ifcase \@ptsize \relax
  \newcommand{\miniscule}{\@setfontsize\miniscule{4}{5}}
\or
  \newcommand{\miniscule}{\@setfontsize\miniscule{5}{6}}
\or
  \newcommand{\miniscule}{\@setfontsize\miniscule{5}{6}}
\fi
\makeatother

\DeclareRobustCommand{\optbar}[1]{\shortstack{{\miniscule (\rule[.5ex]{1.25em}{.18mm})}
  \\ [-.7ex] $#1$}}

\def\mup        {{\ensuremath{\Pmu^+}}\xspace}
\def\mun        {{\ensuremath{\Pmu^-}}\xspace} 

\def\mumu       {{\ensuremath{\Pmu^+\Pmu^-}}\xspace}

\def\squark    {{\ensuremath{\Ps}}\xspace}

\def\pion   {{\ensuremath{\Ppi}}\xspace}

\def\pip    {{\ensuremath{\pion^+}}\xspace}
\def\pim    {{\ensuremath{\pion^-}}\xspace}
\def\pipm   {{\ensuremath{\pion^\pm}}\xspace}

\def\kaon    {{\ensuremath{\PK}}\xspace}

\def\KorKbar {\kern \thebaroffset\optbar{\kern -\thebaroffset \PK}{}\xspace}

\def\Kp      {{\ensuremath{\kaon^+}}\xspace}
\def\Km      {{\ensuremath{\kaon^-}}\xspace}

\def\D       {{\ensuremath{\PD}}\xspace}

\def\DorDbar {\kern \thebaroffset\optbar{\kern -\thebaroffset \PD}\xspace}

\def\Dp      {{\ensuremath{\D^+}}\xspace}
\def\Dm      {{\ensuremath{\D^-}}\xspace}

\def\DpDm    {\ensuremath{\Dp {\kern -0.16em \Dm}}\xspace}

\def\B       {{\ensuremath{\PB}}\xspace}
\def\Bbar    {{\ensuremath{\offsetoverline{\PB}}}\xspace}

\def\BorBbar {\kern \thebaroffset\optbar{\kern -\thebaroffset \PB}\xspace}
\def\Bz      {{\ensuremath{\B^0}}\xspace}

\def\Bd      {{\ensuremath{\B^0}}\xspace}
\def\Bdb     {{\ensuremath{\Bbar{}^0}}\xspace}
\def\BdorBdbar {\kern \thebaroffset\optbar{\kern -\thebaroffset \Bd}\xspace}

\def\Bs      {{\ensuremath{\B^0_\squark}}\xspace}
\def\Bsb     {{\ensuremath{\Bbar{}^0_\squark}}\xspace}
\def\BsorBsbar {\kern \thebaroffset\optbar{\kern -\thebaroffset \Bs}\xspace}

\def\Bds     {{\ensuremath{\B_{(\squark)}^0}}\xspace}
\def\Bdsb    {{\ensuremath{\Bbar{}_{(\squark)}^0}}\xspace}
\def\BdorBs  {\Bds}
\def\BdorBsbar  {\Bdsb}

\def\jpsi     {{\ensuremath{{\PJ\mskip -3mu/\mskip -2mu\Ppsi}}}\xspace}

\def\Y#1S{\ensuremath{\PUpsilon{(#1S)}}\xspace}

\def\LorLbar     {\kern \thebaroffset\optbar{\kern -\thebaroffset \PLambda}\xspace}

\def\to                 {\ensuremath{\rightarrow}\xspace}

\def\CP                {{\ensuremath{C\!P}}\xspace}

\def\AT#1     {\ensuremath{A_{\mathrm{T}}^{#1}}\xspace}           

\def\C#1      {\ensuremath{\mathcal{C}_{#1}}\xspace}                       
\def\Cp#1     {\ensuremath{\mathcal{C}_{#1}^{'}}\xspace}                    
\def\Ceff#1   {\ensuremath{\mathcal{C}_{#1}^{\mathrm{(eff)}}}\xspace}        
\def\Cpeff#1  {\ensuremath{\mathcal{C}_{#1}^{'\mathrm{(eff)}}}\xspace}       
\def\Ope#1    {\ensuremath{\mathcal{O}_{#1}}\xspace}                       
\def\Opep#1   {\ensuremath{\mathcal{O}_{#1}^{'}}\xspace}                    

\newcommand{\aunit}[1]{\ensuremath{\text{\,#1}}}       

\newcommand{\tev}{\aunit{Te\kern -0.1em V}\xspace}
\newcommand{\gev}{\aunit{Ge\kern -0.1em V}\xspace}
\newcommand{\mev}{\aunit{Me\kern -0.1em V}\xspace}
\newcommand{\kev}{\aunit{ke\kern -0.1em V}\xspace}
\newcommand{\ev}{\aunit{e\kern -0.1em V}\xspace}
 
\newcommand{\mevc}{\ensuremath{\aunit{Me\kern -0.1em V\!/}c}\xspace}
\newcommand{\gevc}{\ensuremath{\aunit{Ge\kern -0.1em V\!/}c}\xspace}
\newcommand{\mevcc}{\ensuremath{\aunit{Me\kern -0.1em V\!/}c^2}\xspace}
\newcommand{\gevcc}{\ensuremath{\aunit{Ge\kern -0.1em V\!/}c^2}\xspace}

\def\ps   {\ensuremath{\aunit{ps}}\xspace}

\def\gsim{{~\raise.15em\hbox{$>$}\kern-.85em
          \lower.35em\hbox{$\sim$}~}\xspace}
\def\lsim{{~\raise.15em\hbox{$<$}\kern-.85em
          \lower.35em\hbox{$\sim$}~}\xspace}

\def\tell1  {TELL1\xspace}
\def\ukl1   {UKL1\xspace}

\newcommand{\ie}{\mbox{\itshape i.e.}\xspace}

\newcommand{\lhcborcid}[1]{\href{https://orcid.org/#1}{\hspace*{0.1em}\raisebox{-0.45ex}{\includegraphics[width=1em]{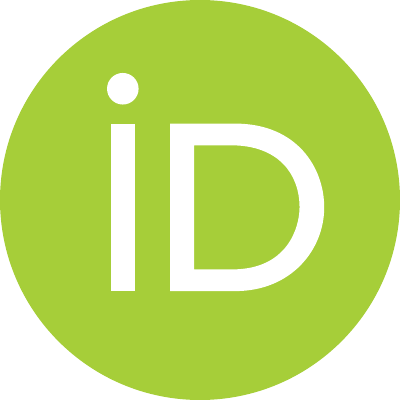}}}}

\usepackage{cite} 
\usepackage{mciteplus}

\usepackage{relsize} 
\usepackage[hang,flushmargin]{footmisc} 
\usepackage{comment}

\def\sintb{\ensuremath{\sin(2\beta)}\xspace}
\def\costb{\ensuremath{\cos(2\beta)}\xspace}

\newcommand{\appropto}{\mathrel{\vcenter{
  \offinterlineskip\halign{\hfil$##$\cr
    \propto\cr\noalign{\kkkkkern2pt}\sim\cr\noalign{\kern-2pt}}}}}

\mathchardef\mhyphen="2D    

\allowdisplaybreaks 

\begin{document}

\renewcommand{\thefootnote}{\fnsymbol{footnote}}
\setcounter{footnote}{1}

\begin{titlepage}
\pagenumbering{roman}

{\bf\boldmath\huge
\begin{center}
\papertitle
\end{center}}

\vspace*{1.5cm}

\begin{center}
Tim Gershon$^1$\lhcborcid{0000-0002-3183-5065}, 
Thomas Latham$^1$\lhcborcid{0000-0002-7195-8537}, 
Peilian Li$^{2,3}$\lhcborcid{0000-0003-2740-9765}, 
Andy Morris$^4$\lhcborcid{0000-0001-6644-9888},\\ 
Wenbin Qian$^3$\lhcborcid{0000-0003-3932-7556},
Mark Whitehead$^5$\lhcborcid{0000-0002-2142-3673}, 
Ao Xu$^{6,7}\lhcborcid{0000-0002-8521-1688}$\bigskip\\
{\normalfont\itshape\footnotesize
$ ^1$University of Warwick, Coventry, United Kingdom\\
$ ^2$European Organization for Nuclear Research (CERN), Geneva, Switzerland\\
$ ^3$University of Chinese Academy of Sciences, Beijing, China\\
$ ^4$Aix Marseille Univ, CNRS/IN2P3, CPPM, Marseille, France\\
$ ^5$School of Physics and Astronomy, University of Glasgow, Glasgow, United Kingdom\\
$ ^6$INFN Sezione di Pisa, Pisa, Italy\\
$ ^7$Scuola Normale Superiore, Pisa, Italy
}
\end{center}

\vspace{\fill}

\begin{abstract}
  \noindent
  Methods have been proposed recently to weight data in order to allow visualisation of \CP-violation effects in transitions of neutral $B$ mesons to multibody final states that are not \CP-eigenstates.
  These are useful since integration of the unweighted data over the phase space would otherwise wash out the effects of interest.  
  A similar method, elaborated upon here, with a different weighting function can also be used to visualise \CP-conserving $\BdorBs\text{--}\BdorBsbar$ oscillations, rather than the \CP-violating asymmetries.
  Together with the other weighting functions, this new method could be useful, for example, to demonstrate the accuracy of the calibration of the flavour tagging algorithms that are crucial for analyses such as the measurement of the \CP-violating phase in $\Bs \to \jpsi\phi$ decays.
  Their application to the formalism in common use for such decays is explicitly demonstrated.
\end{abstract}

\vspace{\fill}

\end{titlepage}

\newpage
\setcounter{page}{2}
\mbox{~}

\clearpage

\renewcommand{\thefootnote}{\arabic{footnote}}
\setcounter{footnote}{0}

\pagestyle{plain} 
\setcounter{page}{1}
\pagenumbering{arabic}

\section{Introduction}
\label{sec:intro}

Decays of neutral \B mesons to multibody self-conjugate final states contain, in general, an admixture of \CP-even and \CP-odd components.
Since \CP-violation effects in the decay-time distributions associated with these components have opposite signs, they tend to be diluted by integrating over the final-state phase space.  
Additionally, effects related to interference between the \CP-even and -odd amplitudes are completely removed by such integration.
For this reason, the authors have recently proposed weighting functions that can be applied to enable visualisation of these \CP-violation effects~\cite{Gershon:2024xkk}.

The $\Bs \to \jpsi\phi$ decay provides an excellent example of a channel where application of this new weighting method would be useful.
This channel is the golden mode to determine the \CP-violating phase $\phi_s$ arising from the interference between $\Bs\text{--}\Bsb$ mixing and $\Bs \to \jpsi\phi$ decay amplitudes~\cite{Dighe:1998vk,Dunietz:2000cr}, with recent results starting to show an indication of a non-zero effect~\cite{LHCb-PAPER-2023-016,ATLAS:2020lbz,CMS:2024znt}.
However, visualisations of the data shown in publications of experimental results to date are not sufficiently powerful for the effect of a non-zero value of $\phi_s$ to be seen by the reader; application of the method of Ref.~\cite{Gershon:2024xkk} would help to rectify this.  
In addition, these analyses also have good sensitivity to the mass difference $\Delta m_s$ between the two physical eigenstates of the $\Bs\text{--}\Bsb$ system, which corresponds to the angular frequency of the oscillations between the two eigenstates.\footnote{Units with $\hbar = c = 1$ are used throughout this document.}
Indeed, the good sensitivity of $\Bs \to \jpsi\phi$ decays to $\Delta m_s$ is known from previous measurements~\cite{LHCb-PAPER-2013-002,LHCb-PAPER-2019-013,CMS:2020efq}, but no visualisation has been provided in any of these works that shows the oscillations.
Visualisation of the \CP-conserving oscillations in $\Bs \to \jpsi\phi$ decays would be especially useful to validate the performance of the flavour tagging algorithms that are necessary to obtain sensitivity to decay-time-dependent asymmetries.

In this paper, a new weighting method is presented that allows visualisation of \CP-conserving oscillations in decays of neutral \B mesons to multibody self-conjugate final states.
It is intended to be used together with the methods from Ref.~\cite{Gershon:2024xkk}, which are shown to also provide sensitivity to \CP-conserving oscillations.
The method is general, and is illustrated with both $\Bd \to D_{\CP} \pip\pim$ decays (as in Ref.~\cite{Gershon:2024xkk}) and $\Bs \to \jpsi \phi$ decays.
For the latter, the formalism is expressed explicitly in terms of the transversity amplitudes of the vector-vector final state, as conventionally used in the literature for this decay mode.  

\section{Method}
\label{sec:methodology}

Using similar notation as Ref.~\cite{Gershon:2024xkk}, and following the conventions of Refs.~\cite{PDG-review,PDG2024}, the decay-time-dependent rates for neutral \B mesons tagged as \BdorBs or \BdorBsbar at $t=0$ are given by 
\begin{multline}
    \Gamma[\BdorBs \to f(t)] \propto e^{-t/\tau}\Big{(}
        \left(|\mathcal{A}_f|^2+|\overline{\mathcal{A}}_f|^2\right) \cosh(\Delta\Gamma t/2)
        +2\mathcal{R}e\Big{(}e^{i\phi}\mathcal{A}_f^*\overline{\mathcal{A}}_f\Big{)} \sinh(\Delta\Gamma t/2)\\
        -2\mathcal{I}m\Big{(}e^{i\phi}\mathcal{A}_f^*\overline{\mathcal{A}}_f\Big{)} \sin(\Delta m t)
        +\left(|\mathcal{A}_f|^2-|\overline{\mathcal{A}}_f|^2\right) \cos(\Delta m t)
    \Big{)}\,,
    \label{eq:B0_to_fCP_wrt_time_ideal}
\end{multline}
\vspace{-5ex}
\begin{multline}
    \Gamma[\BdorBsbar \to f(t)] \propto e^{-t/\tau}\Big{(}
        \left(|\mathcal{A}_f|^2+|\overline{\mathcal{A}}_f|^2\right) \cosh(\Delta\Gamma t/2)
        +2\mathcal{R}e\Big{(}e^{i\phi}\mathcal{A}_f^*\overline{\mathcal{A}}_f\Big{)} \sinh(\Delta\Gamma t/2)\\
        +2\mathcal{I}m\Big{(}e^{i\phi}\mathcal{A}_f^*\overline{\mathcal{A}}_f\Big{)} \sin(\Delta m t)
        -\left(|\mathcal{A}_f|^2-|\overline{\mathcal{A}}_f|^2\right) \cos(\Delta m t)
    \Big{)}\,.
    \label{eq:B0bar_to_fCP_wrt_time_ideal}
\end{multline}
Here, the amplitudes $\mathcal{A}_f$ and $\overline{\mathcal{A}}_f$ are those for $\BdorBs$ and $\BdorBsbar$ decays, respectively, to the point $f$ in the final-state phase space.
It is assumed that these amplitudes each include contributions with only one set of CKM matrix elements so that there is no \CP\ violation in decay, and therefore the \CP-violating phase difference from the interference of mixing and decay $\phi$ can be factored out from the amplitudes.  
(For a \CP-eigenstate final state $f$ with \CP-eigenvalue $\eta_{\CP}$ one has $\eta_{\CP} e^{i\phi} = \frac{q}{p}\frac{\overline{\mathcal{A}}_f}{\mathcal{A}_f}$, where $p$ and $q$ give the admixtures of the \BdorBs\ and \BdorBsbar\ components in the mass eigenstates and the same phase convention as Ref.~\cite{Gershon:2024xkk} is used.\footnote{Elsewhere in the literature the phase of $\frac{q}{p}\frac{\overline{\mathcal{A}}_f}{\mathcal{A}_f}$ is sometimes written as $-\phi$.})
The parameters $\Delta m$ and $\Delta \Gamma$ are the mass and width difference between the heavier and lighter of the two mass eigenstates of the $\BdorBs\text{--}\BdorBsbar$ system,\footnote{Note that with this convention, the value of $\Delta \Gamma$ for the $\Bs\text{--}\Bsb$ system is negative.  The opposite convention with $\Delta \Gamma_s > 0$ is also widely used in the literature.} while $\tau$ is the inverse of the average of the two widths.  
The decay-time $t$ is taken to range over non-negative values, as is the case for $B$ mesons produced in LHC collisions.
The normalisation of the decay rates is omitted from Eqs.~\eqref{eq:B0_to_fCP_wrt_time_ideal} and~\eqref{eq:B0bar_to_fCP_wrt_time_ideal} as it is not relevant for the discussion here.

In Ref.~\cite{Gershon:2024xkk}, weighting functions $w^{\mathcal{I}m}$ and $w^{\mathcal{R}e}$ are defined in order to obtain sensitivity to the two components in the coefficient of the sinusoidal term of Eqs.~\eqref{eq:B0_to_fCP_wrt_time_ideal} and~\eqref{eq:B0bar_to_fCP_wrt_time_ideal},
\begin{equation}
    \label{eq:Im-part}
    {}\mp{}2\mathcal{I}m\Big{(}e^{i\phi}\mathcal{A}_f^*\overline{\mathcal{A}}_f\Big{)} = {}\mp{}2\left(\mathcal{I}m(\mathcal{A}_f^*\overline{\mathcal{A}}_f) \,\cos\phi + \mathcal{R}e(\mathcal{A}_f^*\overline{\mathcal{A}}_f) \,\sin\phi \right) \, .
\end{equation}
This is done by noting that $\mathcal{I}m(\mathcal{A}_f^*\overline{\mathcal{A}}_f)$ has the same magnitude but changes sign between the point $f$ in phase space and the related point obtained when replacing all final-state particles with their antiparticles (for $\Bs \to \jpsi [\to \mup\mun]\phi [\to \Kp\Km]$ decays this means $\mup \leftrightarrow \mun$ and $\Kp \leftrightarrow \Km$), while $\mathcal{R}e(\mathcal{A}_f^*\overline{\mathcal{A}}_f)$ does not change sign.
These terms are therefore referred to as being respectively antisymmetric and symmetric with respect to this transformation. 
Due to these properties, the term involving $\mathcal{I}m(\mathcal{A}_f^*\overline{\mathcal{A}}_f)$ will vanish identically when integrating over the phase space, while that involving $\mathcal{R}e(\mathcal{A}_f^*\overline{\mathcal{A}}_f)$ is likely to be diluted as it can have both positive and negative values.  
To avoid cancellations or dilutions in the first (second) term, the weighting function should have the same symmetry properties and the same sign as $\mathcal{I}m(\mathcal{A}_f^*\overline{\mathcal{A}}_f)$ ($\mathcal{R}e(\mathcal{A}_f^*\overline{\mathcal{A}}_f)$), and should also be constant in the limit of a pure \CP-eigenstate.  
The corresponding functions, which depend on the position in phase-space $f$, are~\cite{Gershon:2024xkk}
\begin{equation}
\label{eq:weighting-functions}
    w^{\mathcal{I}m}_f = \frac{2\,\mathcal{I}m(\mathcal{A}_f^*\overline{\mathcal{A}}_f)}{|\mathcal{A}_f|^2+|\overline{\mathcal{A}}_f|^2}\,,
    \quad\text{and}\quad
    w^{\mathcal{R}e}_f = \frac{2\,\mathcal{R}e(\mathcal{A}_f^*\overline{\mathcal{A}}_f)}{|\mathcal{A}_f|^2+|\overline{\mathcal{A}}_f|^2}\,,
\end{equation}
respectively, where the normalisation is chosen such that these functions are each in the range $\left[-1,+1\right]$.

In a similar way, a weighting function can be chosen to avoid cancellation in the coefficient of the cosine term of Eqs.~\eqref{eq:B0_to_fCP_wrt_time_ideal} and~\eqref{eq:B0bar_to_fCP_wrt_time_ideal} when integrating over the phase space.
This coefficient, ${}\pm\left(|\mathcal{A}_f|^2-|\overline{\mathcal{A}}_f|^2\right)$, is antisymmetric with respect to charge conjugation of the final state (as stated earlier, absence of \CP violation in decay is assumed) and therefore vanishes after integration of the unweighted data.
The $\cos(\Delta m t)$ oscillations, which are not related to \CP violation in the interference between mixing and decay, can be preserved by weighting with 
\begin{equation}
\label{eq:new-weighting-function}
    w^{\rm fs}_f = \frac{|\mathcal{A}_f|^2-|\overline{\mathcal{A}}_f|^2}{|\mathcal{A}_f|^2+|\overline{\mathcal{A}}_f|^2}\,, 
\end{equation}
which is also normalised in the range $\left[-1,+1\right]$ and takes the constant values of 0 for a pure \CP-eigenstate ($\mathcal{A}_f = \pm \overline{\mathcal{A}}_f$) and $\pm 1$ for a flavour-specific final state ($\overline{\mathcal{A}}_f = 0$ or $\mathcal{A}_f = 0$).
The superscript in the notation $w^{\rm fs}_f$ indicates that this weighting function gives a measure of how flavour-specific the position $f$ in the final-state phase space is.

A weighted decay-time asymmetry can be obtained in a similar way as done in Ref.~\cite{Gershon:2024xkk}, where the weighting is applied only to the decay-time distributions in the numerator, and not the denominator,
\begin{equation}
\label{eq:weighted-acpt}
    A^{w\mhyphen{\rm fs}}_{\CP}(t) \equiv \frac{\int_{\rm PS} w^{\rm fs}_f \left( \Gamma[\Bdb \to f(t)] - \Gamma[\Bd \to f(t)]\right) {\rm d}\Omega}{\int_{\rm PS} \left( \Gamma[\Bdb \to f(t)]+\Gamma[\Bd \to f(t)] \right) {\rm d}\Omega}\,.
\end{equation}
where ${\rm d}\Omega$ represents an element in the final-state phase-space (PS).
This can be written as 
\small 
\begin{eqnarray}
\label{eq:wacp_fs_def-Bs}
  A^{w\mhyphen{\rm fs}}_{\CP}(t) & = & S^{w\mhyphen{\rm fs}} \sin(\Delta m t) - C^{w\mhyphen{\rm fs}} \cos(\Delta m t)\,,\\
\label{eq:wS_fs_def-Bs-withDG}
  \!\!\!\!\!\text{where}\ S^{w\mhyphen{\rm fs}} & = & \frac{ 2 \cos \phi \int_{\rm PS}\left(|\mathcal{A}_f|^2-|\overline{\mathcal{A}}_f|^2\right) \mathcal{I}m(\mathcal{A}_f^*\overline{\mathcal{A}}_f)/\left(|\mathcal{A}_f|^2+|\overline{\mathcal{A}}_f|^2\right) {\rm d}\Omega }{\int_{\rm PS}\left(|\mathcal{A}_f|^2+|\overline{\mathcal{A}}_f|^2\right)\cosh(\Delta\Gamma t/2) +2\mathcal{R}e\Big{(}e^{i\phi}\mathcal{A}_f^*\overline{\mathcal{A}}_f\Big{)} \sinh(\Delta\Gamma t/2)\,{\rm d}\Omega} \,,\\
\label{eq:wC_fs_def-Bs-withDG}
  \!\!\!\!\!\text{and}\ C^{w\mhyphen{\rm fs}} & = & \frac{ \int_{\rm PS}\left(|\mathcal{A}_f|^2-|\overline{\mathcal{A}}_f|^2\right)^2 /\left(|\mathcal{A}_f|^2+|\overline{\mathcal{A}}_f|^2\right) {\rm d}\Omega}{\int_{\rm PS}\left(|\mathcal{A}_f|^2+|\overline{\mathcal{A}}_f|^2\right)\cosh(\Delta\Gamma t/2) +2\mathcal{R}e\Big{(}e^{i\phi}\mathcal{A}_f^*\overline{\mathcal{A}}_f\Big{)} \sinh(\Delta\Gamma t/2)\,{\rm d}\Omega}\,.
\end{eqnarray}
\normalsize
Neglecting the impact of a non-zero value of $\Delta \Gamma$ in the denominator, as is a good approximation in the $\Bd\text{--}\Bdb$ system and is also commonly done when making folded asymmetry plots of $\Bs\text{--}\Bsb$ oscillations (see the discussion in Sec.~5 of Ref.~\cite{Gershon:2024xkk}), these simplify to 
\begin{eqnarray}
\label{eq:wS_fs_def-Bs}
  S^{w\mhyphen{\rm fs}} & \approx & \frac{ 2 \cos \phi \int_{\rm PS}\left(|\mathcal{A}_f|^2-|\overline{\mathcal{A}}_f|^2\right) \mathcal{I}m(\mathcal{A}_f^*\overline{\mathcal{A}}_f)/\left(|\mathcal{A}_f|^2+|\overline{\mathcal{A}}_f|^2\right) {\rm d}\Omega }{\int_{\rm PS}\left(|\mathcal{A}_f|^2+|\overline{\mathcal{A}}_f|^2\right)\,{\rm d}\Omega} \,,\\
\label{eq:wC_fs_def-Bs}
  \!\!\!\!\!\text{and}\ C^{w\mhyphen{\rm fs}} & \approx & \frac{ \int_{\rm PS}\left(|\mathcal{A}_f|^2-|\overline{\mathcal{A}}_f|^2\right)^2 /\left(|\mathcal{A}_f|^2+|\overline{\mathcal{A}}_f|^2\right) {\rm d}\Omega}{\int_{\rm PS}\left(|\mathcal{A}_f|^2+|\overline{\mathcal{A}}_f|^2\right)\,{\rm d}\Omega}\,,
\end{eqnarray}
where all terms except $\cos \phi$ can be calculated within a given amplitude model (\ie\ if $\mathcal{A}_f$ and $\overline{\mathcal{A}}_f$ are known).
In the case that $|C^{w\mhyphen{\rm fs}}| \gg |S^{w\mhyphen{\rm fs}}|$, the weighted asymmetry $A^{w\mhyphen{\rm fs}}_{\CP}(t)$ will primarily visualise the \CP-conserving cosine oscillations.  
This is generically expected, since the integrand in the numerator of the definition of $S^{w\mhyphen{\rm fs}}$ can take both positive and negative values, and therefore some cancellation will occur on integration, while that in the definition of $C^{w\mhyphen{\rm fs}}$ is positive definite. 
Note that apart from a multiplicative factor of $\cos\phi$, $S^{w\mhyphen{\rm fs}}$ is identical to the factor $C^{w\mhyphen\mathcal{I}m}$ that appears in one of the weighted decay-time asymmetries discussed in Ref.~\cite{Gershon:2024xkk}.

\section{Illustrations of the method}
\label{sec:illustration}

\subsection{\texorpdfstring{\boldmath $\Bd \to D_{\CP} \pip\pim$}{B0 -> D\_CP pi+pi-}}
\label{sec:illustration:Dpipi}

The potential of the method can be illustrated with the $\Bz \to D_{\CP}\pip\pim$ decay, as done in Ref.~\cite{Gershon:2024xkk}.
The amplitude model is taken from the isobar model results of Ref.~\cite{LHCb-PAPER-2014-070}, implemented in the {\sc Laura}$^{++}$ software package~\cite{Back:2017zqt}.
The largest components in the model are $D\rho^0$, $D(\pip\pim)_{\rm S\,wave}$ (including $Df_0(500)$) and $Df_2(1270)$ together with $D_2^*(2460)^-\pip$, $D_0^*(2400)^-\pip$ and a $(D\pim)_{\rm P\,wave}\pip$ term.
The $D^*(2010)^-\pip$ component is vetoed with the requirement $m(D\pipm) > 2.1 \gev$, since it is too narrow to contribute significant interference effects that would enhance the sensitivity to \CP violation.
(Inclusion of the $D^*(2010)^-\pip$ component, which is treated as flavour-specific in this study, would enhance the sensitivity to $\Delta m_d$ but the selection is kept consistent with that of Ref.~\cite{Gershon:2024xkk}.)
The hadronic factors of Eqs.~\eqref{eq:wS_fs_def-Bs} and~\eqref{eq:wC_fs_def-Bs} are evaluated with this model and are found to be 
\begin{equation}
    S^{w\mhyphen{\rm fs}} = -0.002 \times \costb
    \quad\text{and}\quad  
    C^{w\mhyphen{\rm fs}} = 0.504\,,
\end{equation}
where for the $\Bz \to D_{\CP}\pip\pim$ decay the weak phase difference between decays with and without mixing is $\phi = -2\beta$ to an excellent approximation.
As noted previously, the hadronic factor entering $S^{w\mhyphen{\rm fs}}$ is identical to that given in Eq.~(23) of Ref.~\cite{Gershon:2024xkk}.

\begin{figure}[!tb]
    \centering
    \includegraphics[width=0.48\linewidth]{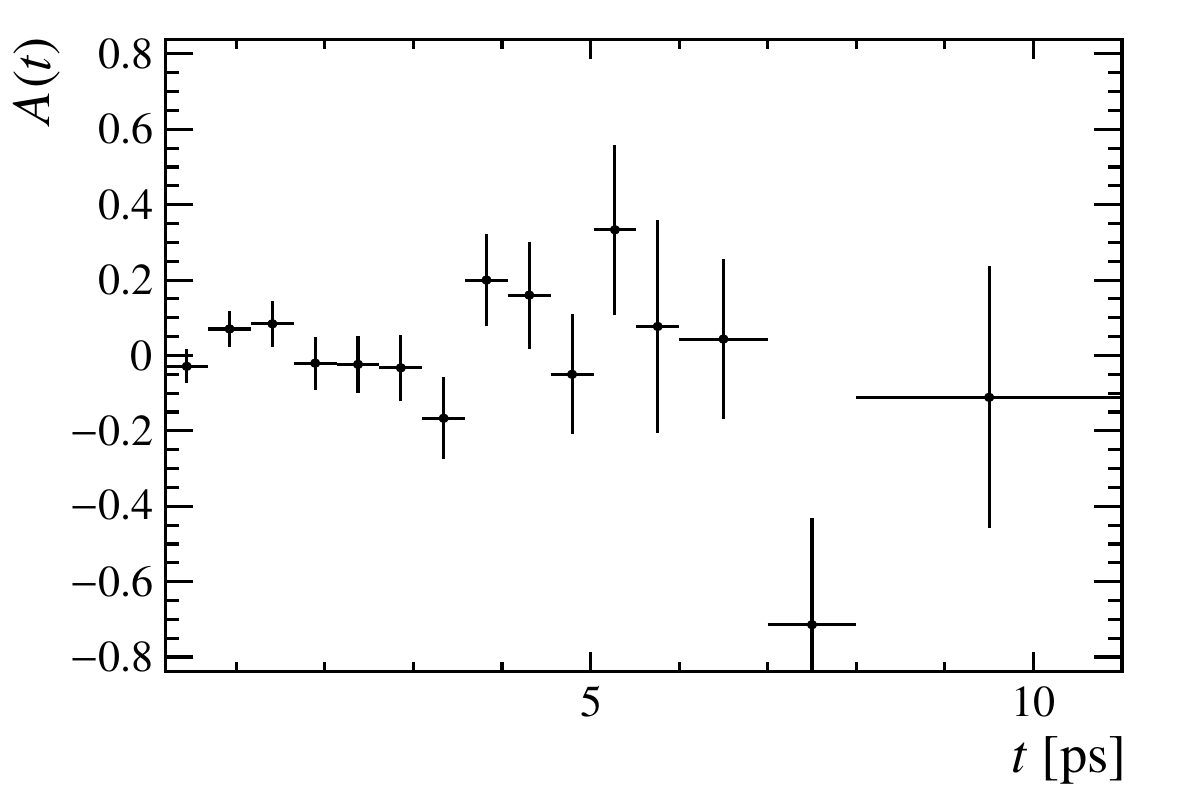}
    \includegraphics[width=0.48\linewidth]{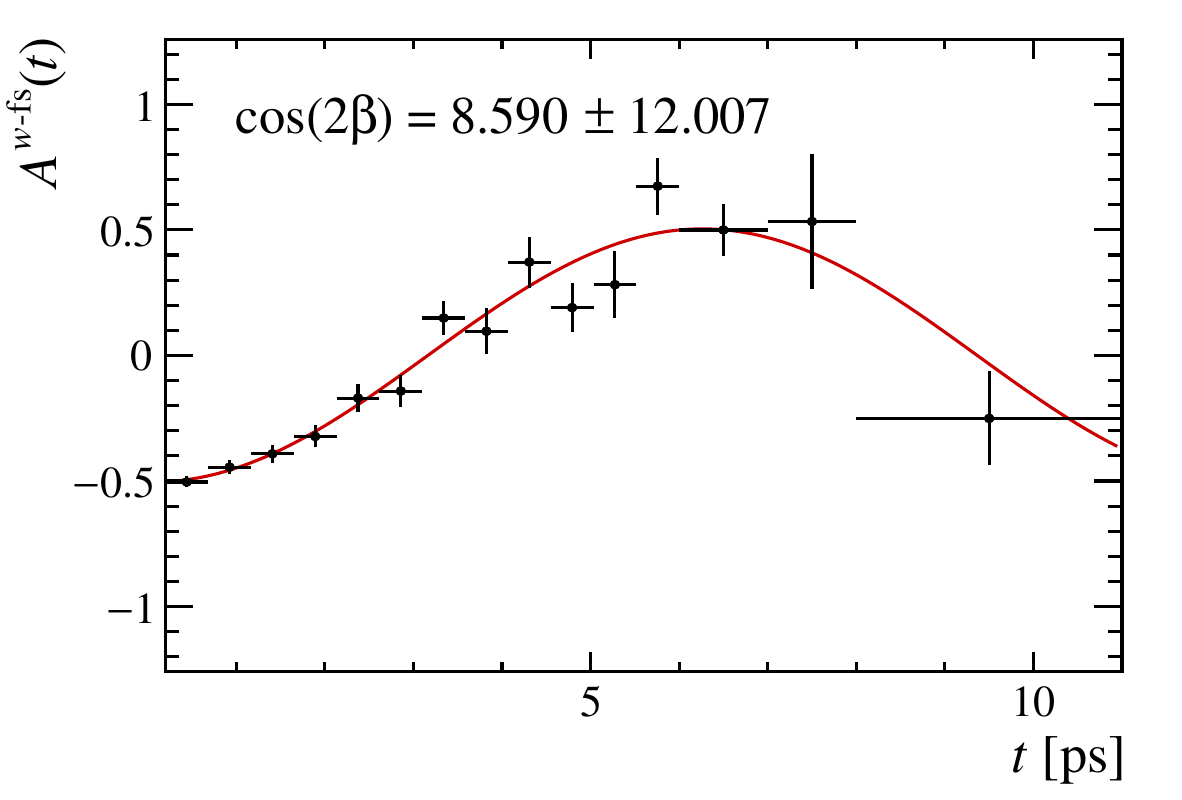}
    \caption{Distributions of time-dependent asymmetry in simulated $\Bz \to D_{\CP}\pip\pim$ decays, with the time-dependent decay rates in the numerator (left)~unweighted, (right)~weighted by $w^{\rm fs}_f$.
    The statistical uncertainties are evaluated with a bootstrap method~\cite{efron:1979}. 
    Results of the fits described in the text are also shown.
    The left plot is identical to Fig.~4~(top) from Ref.~\cite{Gershon:2024xkk}.
    }
    \label{fig:wacp_toy}
\end{figure}

The same pseudoexperiments as in Ref.~\cite{Gershon:2024xkk} are used to illustrate the impact of weighting with the $w^{\rm fs}$ function.
Simulated decays are generated with the amplitude model described above, the values of $\tau_{\Bz}$, $\Delta m_d$, $\sintb$ and $\costb$ fixed to their world averages~\cite{HFLAV24}, and $\Delta \Gamma_d = 0$. 
The sample size corresponds approximately to the expected equivalent yield of perfectly tagged $\Bz \to D_{\CP}\pip\pim$ decays available in the Run~1 and~2 LHCb data sample.
Experimental effects such as backgrounds, imperfect flavour tagging and dependence of the efficiency on decay-time and position in the phase space are not included (the impact of experimental effects is discussed in Sec.~4 of Ref.~\cite{Gershon:2024xkk}).  
The weighted decay-time asymmetry of Eq.~\eqref{eq:wacp_fs_def-Bs} is compared to the unweighted decay-time-asymmetry in Fig.~\ref{fig:wacp_toy}; the impact of the weighting is evident and allows the $\Bd\text{--}\Bdb$ oscillations to be clearly seen.
The weighted decay-time-asymmetry is fitted with $\costb$ as a free parameter and all hadronic factors fixed.
As expected, this distribution has negligible sensitivity to $\costb$ due to the small value of the hadronic factor entering $S^{w\mhyphen{\rm fs}}$. 

\subsection{\texorpdfstring{\boldmath $\Bs \to \jpsi \phi$}{Bs -> J/psi phi}}
\label{sec:illustration:Jpsiphi}

The $\Bz \to D_{\CP}\pip\pim$ decay model described in the previous subsection includes flavour-specific amplitudes such as that for $\Bz \to D_2^*(2460)^-\pip$.
It is therefore of interest to consider also a neutral $B$ meson decay to a multibody final state that does not include any manifestly flavour-specific amplitude.  
The $\Bs \to \jpsi \phi$ decay, where subsequent $\jpsi \to \mumu$ and $\phi(1020) \to \Kp\Km$ decays are implied, is ideal for this purpose, bearing in mind that it is also one of the channels where the impact of the weighting method may be most powerful. 
Since this channel was not used to illustrate the method in Ref.~\cite{Gershon:2024xkk}, the effects of all three weighting functions $w^{\mathcal{I}m}$, $w^{\mathcal{R}e}$ and $w^{\rm fs}$ are discussed here.

\subsubsection{Formalism for \texorpdfstring{\boldmath $\Bs \to \jpsi \phi$}{Bs -> J/psi phi} decays}
\label{sec:illustration:Jpsiphi:formalism}

The decay-time-dependent rate for a neutral \BdorBs meson transition to two vector resonances, such as $\Bs \to \jpsi\phi$, can be expressed in terms of the total amplitudes for \BdorBs and \BdorBsbar decays to each position in the final state phase-space, in the same way as done in Eqs.~\eqref{eq:B0_to_fCP_wrt_time_ideal} and~\eqref{eq:B0bar_to_fCP_wrt_time_ideal}.
However, the expressions are usually given in a different way in the literature, so it is relevant to demonstrate the correspondence between the different formalisms used.  
Following Ref.~\cite{LHCb-PAPER-2013-002}, the total amplitude is written as the sum of contributions from three polarisation states, in the transversity basis, plus a contribution from a non-$\phi$ $\Kp\Km$ ${\rm S}$-wave term which has been found experimentally to be non-negligible in the $\phi$ mass window.  
Each of these is written as a constant complex coefficient ($A_0$, $A_\perp$, $A_\parallel$, $A_{\rm S}$) multiplied by a factor that describes the variation of each component over the phase space.
This sum of products of constant coefficients with factors containing phase-space variation is similar to what is done in Dalitz-plot analyses such as for $\Bz \to D_{\CP}\pip\pim$, except that the sum is over polarisation amplitudes rather than resonances, and the phase space is described by decay angles rather than Dalitz-plot co-ordinates.
The phases of the amplitudes are denoted $\delta_0$, $\delta_\perp$, $\delta_\parallel$ and $\delta_{\rm S}$. 
The normalisation $|A_0|^2 + |A_\perp|^2 + |A_\parallel|^2 = 1$ is commonly used; this choice does not affect the discussion here.
For the $\Bs \to \jpsi\phi$ decay, the amplitudes represented by $A_0$ and $A_\parallel$ are \CP-even, while those given by $A_\perp$ and $A_{\rm S}$ are \CP-odd.
In the absence of \CP-violation in decay, each of the complex coefficients $A_0$, $A_\perp$, $A_\parallel$ and $A_{\rm S}$ is the same for \Bs\ and \Bsb\ decay, apart from a common weak phase factor from the interference between mixing and decay, which is factored out.
Following Ref.~\cite{LHCb-PAPER-2013-002} this weak phase factor is denoted $\phi_s$ and has the opposite sign to $\phi$ used in Eqs.~\eqref{eq:B0_to_fCP_wrt_time_ideal} and~\eqref{eq:B0bar_to_fCP_wrt_time_ideal}.
Similarly the convention that $\Delta \Gamma_s >0$ is used in Ref.~\cite{LHCb-PAPER-2013-002} and in the following; this is the opposite sign to $\Delta \Gamma$ in Eqs.~\eqref{eq:B0_to_fCP_wrt_time_ideal} and~\eqref{eq:B0bar_to_fCP_wrt_time_ideal}.

Where the treatment of decays to two vector resonances differs from that of Dalitz-plot analyses is that the decay-time-dependent distributions are commonly written in terms of bilinear combinations of the $A_0$, $A_\perp$, $A_\parallel$ and $A_{\rm S}$ coefficients, rather than in terms of the total amplitudes $\mathcal{A}_f$ and $\overline{\mathcal{A}}_f$ as in Eqs.~\eqref{eq:B0_to_fCP_wrt_time_ideal} and~\eqref{eq:B0bar_to_fCP_wrt_time_ideal}.
The corresponding expressions, modified from Eqs.~(1)--(2) of Ref.~\cite{LHCb-PAPER-2013-002} (in turn based on expressions in Refs.~\cite{Dunietz:2000cr,Xie:2009fs}), are
\begin{equation}
    \label{eq:BstoJpsiphi-masterEq}
    \Gamma[\BsorBsbar\to f(t)] \propto e^{-t/\tau_{\Bs}} \sum_{k=1}^{10} N_k h_k(t|\BsorBsbar) f_k(\Omega)\,,
\end{equation}
where
\begin{eqnarray}
    \label{eq:BstoJpsiphi-hk-Bs}
    \!\!\!h_k(t|\Bs) & = &  a_k \cosh(\Delta\Gamma_s t/2) + b_k \sinh(\Delta\Gamma_s t/2) + c_k \cos(\Delta m_s t) + d_k \sin(\Delta m_s t)\,,\\
    \label{eq:BstoJpsiphi-hk-Bsb}
    \!\!\!h_k(t|\Bsb) & = & a_k \cosh(\Delta\Gamma_s t/2) + b_k \sinh(\Delta\Gamma_s t/2) - c_k \cos(\Delta m_s t) - d_k \sin(\Delta m_s t)\,,
\end{eqnarray}
and $N_k$, $a_k$, $b_k$, $c_k$ and $d_k$ are given in Table~\ref{tab:functions} in terms of bilinear combinations of the $A_0$, $A_\perp$, $A_\parallel$ and $A_{\rm S}$ amplitudes and the \CP-violating phase $\phi_s$.\footnote{
    The expressions of Table~\ref{tab:functions} are not generic for all $\BdorBs \to V_1V_2$ decays, where $V_{1,2}$ are vector mesons, but only those where one vector meson decays to a lepton-antilepton pair and the other decays to a meson-antimeson pair.
    This is the relevant case for $\Bs \to \jpsi (\to \mumu) \phi (\to \Kp\Km)$ decays.
    Similar expressions can, however, be obtained for other \B decays to multibody self-conjugate final states through two resonances of any spin and any final state (though if one of the spins is zero there is only one amplitude, the final state is a \CP\ eigenstate, and the formalism discussed here is not necessary).
}
Dependence of the amplitudes on the $\Kp\Km$ invariant mass is integrated over, introducing an additional coherence factor $C_{\rm SP}$ that multiplies $N_8$, $N_9$ and $N_{10}$, but which can be --- and is --- neglected here. 

\begin{table}[!tb]
\caption{\small 
    Definition of angular and decay-time-dependent functions in $\Bs \to \jpsi\phi$ decays assuming absence of \CP~violation in decay, where $S = -\sin(\phi_s)$ and $D = -\cos(\phi_s)$.
    Modified from Ref.~\cite{LHCb-PAPER-2013-002}.
    }
\label{tab:functions}
\scriptsize
\newcommand{\Dterm}{D}
\newcommand{\Sterm}{S}
\newcommand{\thetamu}{\theta_\mu}
\newcommand{\thetaK}{\theta_K}
\newcommand{\phihel}{\phi_{\rm hel}}
\newcommand{\delperp}{\delta_\perp}
\newcommand{\delpar}{\delta_\parallel}
\newcommand{\delzero}{\delta_0}
\newcommand{\dels}{\delta_{\rm S}}
\[
\begin{array}{c|c|c|c|c|c|c}
  k  & f_k(\thetamu,\thetaK, \phihel) & N_k                & a_k                  & b_k & c_k & d_k \\
    \hline
  \rule{0mm}{3mm} 1  & 2\cos^2\thetaK \sin^2\thetamu  						& |A_0|^2         		& 1                         		 		& \Dterm 						& 0 				& -\Sterm \\
  2  &  \sin^2\thetaK \left(1 - \sin^2\thetamu \cos^2\phihel\right)  	& |A_\||^2         		& 1                          				& \Dterm 						& 0 				& -\Sterm \\
  3  &  \sin^2\thetaK \left(1 - \sin^2\thetamu \sin^2\phihel\right)  		& |A_\perp|^2      		& 1                          				& -\Dterm 						& 0 				& \Sterm  \\
  4  & \sin^2\thetaK \sin^2\thetamu \sin2\phihel                     		& |A_\|A_\perp| 	& 0  	& \Sterm\cos(\delperp-\delpar) 		& \sin(\delperp-\delpar) 	& \Dterm \cos(\delperp-\delpar) \\
  5  & \tfrac{1}{2}\sqrt{2}\sin2\thetaK \sin2\thetamu \cos\phihel 		& |A_0 A_\||   		& \cos(\delpar - \delzero)   		&  \Dterm\cos(\delpar - \delzero) 	& 0	& -\Sterm\cos(\delpar - \delzero)  \\
  6  & -\frac{1}{2}\sqrt{2} \sin2\thetaK \sin2\thetamu \sin\phihel 		& |A_0 A_\perp| 	& 0	& \Sterm\cos(\delperp - \delzero)  	& \sin(\delperp - \delzero) &  \Dterm\cos(\delperp - \delzero) \\
  7  &  \tfrac{2}{3}\sin^2\thetamu            						& |A_{\rm S}|^2         		& 1                          				& -\Dterm 						& 0				& \Sterm  \\
  8  & \tfrac{1}{3}\sqrt{6} \sin\thetaK \sin2\thetamu \cos\phihel 		& |A_{\rm S} A_\||   		& 0   	& \Sterm \sin(\delpar - \dels) 		& \cos(\delpar - \dels) 	& \Dterm \sin(\delpar - \dels)  \\
  9  & -\tfrac{1}{3}\sqrt{6}\sin\thetaK \sin2\thetamu \sin\phihel 		& |A_{\rm S} A_\perp| 	& \sin(\delperp - \dels) 			& -\Dterm \sin(\delperp - \dels) 		& 0	& \Sterm\sin(\delperp - \dels) \\
  10 & \tfrac{4}{3}\sqrt{3}\cos\thetaK \sin^2\thetamu 				& |A_{\rm S} A_0|    		& 0   	& \Sterm\sin(\delzero - \dels) 		& \cos(\delzero - \dels) 	&  \Dterm\sin(\delzero - \dels) \\
  
\end{array}
\]
\end{table}

The $f_k(\Omega)$ functions are also given in Table~\ref{tab:functions}, where the phase space $\Omega$ is expressed in terms of three decay angles: the polar angle $\theta_K$ is defined as the angle between the $\Kp$ momentum and the direction opposite to the \Bs\ momentum in the $\Kp\Km$ centre-of-mass system; the polar angle $\theta_\mu$ is similarly the angle between the $\mup$ momentum and the direction opposite to the \Bs\ momentum in the $\mumu$ centre-of-mass system; the azimuthal angle $\phi_{\rm hel}$ is the angle between the $\Kp\Km$ and $\mumu$ decay planes.
Charge conjugation of the final-state particles corresponds to $\theta_K \leftrightarrow \pi -\theta_K$, $\theta_\mu \leftrightarrow \pi - \theta_\mu$, $\phi_{\rm hel} \leftrightarrow -\phi_{\rm hel}$.
It is relevant to separate the terms with different values of the index $k$ in categories depending on whether the corresponding $f_k(\Omega)$ function is symmetric or asymmetric under charge conjugation.
Terms with $k = 1,2,3,5,7$ or $9$ are symmetric and will be denoted by $k \in {\rm sym}$.
Terms with $k=4,6,8$ or $10$ are asymmetric and will be denoted by $k \in {\rm asym}$ --- by construction, these are the terms that involve interference between \CP-even and \CP-odd amplitudes.

The explicit relations between the total amplitudes $\mathcal{A}_f$ and $\overline{\mathcal{A}}_f$ and the bilinear coefficients $N_k$, $a_k$, $b_k$, $c_k$ and $d_k$ and $f_k(\Omega)$ functions are
\begin{equation}\label{eq:dictionary}
\begin{array}{ccccc}
    |\mathcal{A}_f|^2+|\overline{\mathcal{A}}_f|^2 & = &
    \phantom{-}\sum_{k=1}^{10} N_k a_k f_k(\Omega) & = & 
    \phantom{-}\sum_{k\in{\rm sym}} N_k a_k f_k(\Omega) \,, \\
    2\,\mathcal{R}e\Big{(}\mathcal{A}_f^*\overline{\mathcal{A}}_f\Big{)} & = &
    \phantom{-}\sum_{k\in{\rm sym}} N_k \frac{b_k}{D} f_k(\Omega) & = & 
    - \sum_{k\in{\rm sym}} N_k \frac{d_k}{S} f_k(\Omega) \,, \\
    2\,\mathcal{I}m\Big{(}\mathcal{A}_f^*\overline{\mathcal{A}}_f\Big{)} & = & 
    \phantom{-}\sum_{k\in{\rm asym}} N_k \frac{b_k}{S} f_k(\Omega) & = & 
    \phantom{-}\sum_{k\in{\rm asym}} N_k \frac{d_k}{D} f_k(\Omega) \,, \\
    |\mathcal{A}_f|^2-|\overline{\mathcal{A}}_f|^2 & = &
    \phantom{-}\sum_{k=1}^{10} N_k c_k f_k(\Omega) & = & 
    \phantom{-}\sum_{k\in{\rm asym}} N_k c_k f_k(\Omega) \,.
\end{array}
\end{equation}
Only terms with $k \in {\rm sym}$ have $a_k \neq 0$ and contribute to the sum in the first line; similarly only terms with $k \in {\rm asym}$ have $c_k \neq 0$ and contribute to the sum in the last line.
The $b_k$ and $d_k$ coefficients are non-zero, in general, for all $k$, and satisfy $\frac{b_k}{D} = -\frac{d_k}{S}$ for $k \in {\rm sym}$ and $\frac{b_k}{S} = \frac{d_k}{D}$ for $k \in {\rm asym}$.

Finally, it should be stressed that the formalism here, as in Ref.~\cite{Gershon:2024xkk}, assumes absence of \CP~violation in decay.
The most recent LHCb analyses of $\Bs \to \jpsi\phi$ decays~\cite{LHCb-PAPER-2019-013,LHCb-PAPER-2023-016} allow generically for \CP~violation in decay, which implies that each transversity amplitude can have a different weak phase factor and modifies the $a_k$, $b_k$, $c_k$ and $d_k$ coefficients from the values given in Table~\ref{tab:functions}.
The formalism used in Refs.~\cite{LHCb-PAPER-2013-002,CMS:2024znt} assumes that any \CP~violation in decay is the same for all amplitudes, so that a common weak phase difference can still be factored out resulting in only minor modifications to the expressions of Table~\ref{tab:functions}.

\subsubsection{Expressions for the weighting functions and weighted decay-time asymmetries}
\label{sec:illustration:Jpsiphi:weights}

Making use of Eq.~\eqref{eq:dictionary} to convert between notations, the weighting function $w^{\rm fs}$, previously defined in Eq.~\eqref{eq:new-weighting-function}, can be written as
\begin{equation}
    \label{eq:wfs-def-Jpsiphi}
    w^{\rm fs}(\Omega) = 
    \frac{\sum_{k=1}^{10} N_k c_k f_k(\Omega)}{\sum_{k=1}^{10} N_k a_k f_k(\Omega)} = 
    \frac{\sum_{k\in {\rm asym}} N_k c_k f_k(\Omega)}{\sum_{k \in {\rm sym}} N_k a_k f_k(\Omega)}\,,
\end{equation}
where the dependence on $\Omega$ has been retained explicitly for consistency.
The weighting functions $w^{\mathcal{I}m}$ and $w^{\mathcal{R}e}$ can also be expressed in this formalism as 
\begin{eqnarray}
    \label{eq:wImRe-defs-Jpsiphi}
    w^{\mathcal{I}m}(\Omega) & = & 
    \frac{\sum_{k\in {\rm asym}}N_k \frac{b_k}{S} f_k(\Omega)}{\sum_{k \in {\rm sym}} N_k a_k f_k(\Omega)} ~ = ~ 
    \frac{\sum_{k\in {\rm asym}}N_k \frac{d_k}{D} f_k(\Omega)}{\sum_{k \in {\rm sym}} N_k a_k f_k(\Omega)}  \,,\\
    w^{\mathcal{R}e}(\Omega) & = &
    \frac{\sum_{k\in {\rm sym}}N_k \frac{b_k}{D} f_k(\Omega)}{\sum_{k \in {\rm sym}} N_k a_k f_k(\Omega)} ~ = ~ 
    \frac{\sum_{k\in {\rm sym}}N_k \left(\frac{-d_k}{S}\right) f_k(\Omega)}{\sum_{k \in {\rm sym}} N_k a_k f_k(\Omega)} \,.
\end{eqnarray}

The $w^{\mathcal{R}e}$ weighted asymmetry by construction gives good sensitivity to \CP~violation, \ie\ to the value of $\sin(\phi_s)$, as will be shown explicitly below.\footnote{A similar weighting function to $w^{\mathcal{R}e}$ has been used to produce Fig.~5 in the supplementary material to Ref.~\cite{LHCb-PAPER-2019-013}, however in that case the sum is only over $k\in(1,2,3,7)$ rather than $k\in{\rm sym}$.} 
The $w^{\mathcal{I}m}$ weighted asymmetry is constructed to give good sensitivity to $\cos(\phi_s)$, but as pointed out in Ref.~\cite{LHCb-PAPER-2013-002} the sinusoidal $d_k$ terms that are not suppressed by the small magnitude of $S$, \ie\ those with $k \in {\rm asym}$, 
offer good sensitivity to the oscillation frequency $\Delta m_s$.
Thus the $w^{\mathcal{I}m}$ weighted asymmetry is also useful to visualise $\Bs$--$\Bsb$ oscillations.
Additionally, depending on the relative values of the phase differences, the $w^{\rm fs}$ weighted asymmetry can also have good sensitivity to the value of $\Delta m_s$.

The weighted decay-time asymmetries can also be written in terms of the bilinear coefficients of Table~\ref{tab:functions} instead of the total amplitudes $\mathcal{A}_f$ and $\overline{\mathcal{A}}_f$; as will be seen this is useful for evaluation of the hadronic coefficients. 
Specifically, the $S^{w\mhyphen{\rm fs}}$ and $C^{w\mhyphen{\rm fs}}$ terms of Eq.~\eqref{eq:wacp_fs_def-Bs} are 
\begin{equation}\label{eq:S-wfs-fullDef}
    S^{w\mhyphen{\rm fs}} = 
    \frac{
        - \mathlarger{\int}_{\rm PS} \frac{\sum_{{k^\prime}\in {\rm asym}} N_{k^\prime} c_{k^\prime} f_{k^\prime}(\Omega)}{\sum_{{k^\prime} \in {\rm sym}} N_{k^\prime} a_{k^\prime} f_{k^\prime}(\Omega)} \sum_{k=1}^{10} N_k d_k f_k(\Omega) \,{\rm d}\Omega
    }{
        \int_{\rm PS} \sum_{k=1}^{10} N_k \left[ a_k \cosh\left(\frac{\Delta\Gamma_s t}{2}\right) + b_k \sinh\left(\frac{\Delta\Gamma_s t}{2}\right) \right] f_k(\Omega)\,{\rm d}\Omega
    }\,,
\end{equation}
and 
\begin{equation}\label{eq:C-wfs-fullDef}
    C^{w\mhyphen{\rm fs}} = 
    \frac{
        \mathlarger{\int_{\rm PS}} \frac{\sum_{{k^\prime}\in {\rm asym}} N_{k^\prime} c_{k^\prime} f_{k^\prime}(\Omega)}{\sum_{{k^\prime} \in {\rm sym}} N_{k^\prime} a_{k^\prime} f_{k^\prime}(\Omega)} \sum_{k=1}^{10} N_k c_k f_k(\Omega) \,{\rm d}\Omega
    }{
        \int_{\rm PS} \sum_{k=1}^{10} N_k \left[ a_k \cosh\left(\frac{\Delta\Gamma_s t}{2}\right) + b_k \sinh\left(\frac{\Delta\Gamma_s t}{2}\right) \right] f_k(\Omega)\,{\rm d}\Omega    
    }\,.
\end{equation}
Noting that $\int_{\rm PS}f_k(\Omega)\,{\rm d}\Omega = \frac{32\pi}{9}$ for $k = 1,2,3,7$ and zero otherwise~\cite{Dighe:1995pd}, the denominator of these expressions can be replaced by 
\begin{equation}
\begin{split}    
    \frac{32\pi}{9} & \left( |A_0|^2 + |A_\parallel|^2 + |A_\perp|^2 + |A_{\rm S}|^2 \right) \cosh\left(\frac{\Delta\Gamma_s t}{2}\right) - {} \\
    & \hspace{3cm} \frac{32\pi}{9} \cos(\phi_s) \left( |A_0|^2 + |A_\parallel|^2 - |A_\perp|^2 - |A_{\rm S}|^2 \right) \sinh\left(\frac{\Delta\Gamma_s t}{2}\right)\,,
\end{split}
\end{equation}
or, taking the limit $\Delta\Gamma_s \to 0$, by
\begin{equation}
    \frac{32\pi}{9} \left( |A_0|^2 + |A_\parallel|^2 + |A_\perp|^2 + |A_{\rm S}|^2 \right) \,.
\end{equation}
The numerator can also be simplified by noting that any function that is asymmetric with respect to charge conjugation of the final-state particles will vanish on integration over the phase space.  
Therefore, since the weighting function $w^{\rm fs}$ is asymmetric, only terms with $k \in {\rm asym}$ need to be considered in the sum over $k$ in the numerator.  
The remaining integrals are non-trivial and are expected to be evaluated numerically.

With these simplifications, the $S^{w\mhyphen{\rm fs}}$ and $C^{w\mhyphen{\rm fs}}$ terms can be written
\begin{eqnarray}
    \label{eq:wS_fs-bilinear}
    S^{w\mhyphen{\rm fs}} & \approx & 
    \frac{
        \cos(\phi_s) \mathlarger{\int}_{\rm PS} \frac{\frac{1}{D}\sum_{k,{k^\prime}\in {\rm asym}} N_k N_{k^\prime} d_k c_{k^\prime} f_k(\Omega) f_{k^\prime}(\Omega)}{\sum_{{k^\prime} \in {\rm sym}} N_{k^\prime} a_{k^\prime} f_{k^\prime}(\Omega)}\,{\rm d}\Omega
    }{
        \frac{32\pi}{9} \left( |A_0|^2 + |A_\parallel|^2 + |A_\perp|^2 + |A_{\rm S}|^2 \right)    
        }\,, \\
    \label{eq:wC_fs-bilinear}
    \text{and}\quad C^{w\mhyphen{\rm fs}} & \approx &
    \frac{
        \mathlarger{\int}_{\rm PS} \frac{\sum_{k,{k^\prime}\in {\rm asym}} N_k N_{k^\prime} c_k c_{k^\prime} f_k(\Omega) f_{k^\prime}(\Omega)}{\sum_{{k^\prime} \in {\rm sym}} N_{k^\prime} a_{k^\prime} f_{k^\prime}(\Omega)}\,{\rm d}\Omega
    }{
        \frac{32\pi}{9} \left( |A_0|^2 + |A_\parallel|^2 + |A_\perp|^2 + |A_{\rm S}|^2 \right)
    }\,.
\end{eqnarray}
where $D = -\cos(\phi_s)$ has been factored out in Eq.~\eqref{eq:wS_fs-bilinear} for consistency with Eqs.~\eqref{eq:wS_fs_def-Bs-withDG} and~\eqref{eq:wS_fs_def-Bs}.

Similarly, the $S^{w\mhyphen\mathcal{R}e}$ term of Eq.~(16) of Ref.~\cite{Gershon:2024xkk} and the $S^{w\mhyphen\mathcal{I}m}$ and $C^{w\mhyphen\mathcal{I}m}$ terms of Eq.~(18) of Ref.~\cite{Gershon:2024xkk} are, neglecting the effect of non-zero $\Delta\Gamma_s$, given by 
\begin{eqnarray}
    \label{eq:wS_Re-bilinear}
    S^{w\mhyphen\mathcal{R}e} & \approx & 
    \frac{
        - \sin(\phi_s) \int_{\rm PS} \frac{\frac{1}{S^2}\sum_{k,{k^\prime}\in {\rm sym}} N_k N_{k^\prime} d_k d_{k^\prime} f_k(\Omega) f_{k^\prime}(\Omega)}{\sum_{{k^\prime} \in {\rm sym}} N_{k^\prime} a_{k^\prime} f_{k^\prime}(\Omega)}\,{\rm d}\Omega    
    }{
        \frac{32\pi}{9} \left( |A_0|^2 + |A_\parallel|^2 + |A_\perp|^2 + |A_{\rm S}|^2 \right)
    } \, \\
    \label{eq:wS_Im-bilinear}
    S^{w\mhyphen\mathcal{I}m} & \approx & 
    \frac{
        \cos(\phi_s) \int_{\rm PS} \frac{\frac{1}{D^2}\sum_{k,{k^\prime}\in {\rm asym}} N_k N_{k^\prime} d_k d_{k^\prime} f_k(\Omega) f_{k^\prime}(\Omega)}{\sum_{{k^\prime} \in {\rm sym}} N_{k^\prime} a_{k^\prime} f_{k^\prime}(\Omega)}\,{\rm d}\Omega    
    }{
        \frac{32\pi}{9} \left( |A_0|^2 + |A_\parallel|^2 + |A_\perp|^2 + |A_{\rm S}|^2 \right)
    } \,, \\
    \label{eq:wC_Im-bilinear}
    \text{and}\quad C^{w\mhyphen\mathcal{I}m} & \approx & 
    \frac{
        \int_{\rm PS} \frac{\frac{1}{D}\sum_{k,{k^\prime}\in {\rm asym}} N_k N_{k^\prime} c_k d_{k^\prime} f_k(\Omega) f_{k^\prime}(\Omega)}{\sum_{{k^\prime} \in {\rm sym}} N_{k^\prime} a_{k^\prime} f_{k^\prime}(\Omega)}\,{\rm d}\Omega    
    }{
        \frac{32\pi}{9} \left( |A_0|^2 + |A_\parallel|^2 + |A_\perp|^2 + |A_{\rm S}|^2 \right)
    } \,,
\end{eqnarray}
where $S = -\sin(\phi_s)$ has been factored out in Eq.~\eqref{eq:wS_Re-bilinear} and $D = -\cos(\phi_s)$ has been factored out in Eq.~\eqref{eq:wS_Im-bilinear} for consistency with Eqs.~(17) and~(19) of Ref.~\cite{Gershon:2024xkk}, respectively.
As noted previously, $S^{w\mhyphen{\rm fs}} = \cos(\phi_s)\,C^{w\mhyphen\mathcal{I}m}$.

\subsubsection{Illustration with simulation}
\label{sec:illustration:Jpsiphi:toy}

\begin{figure}[!tb]
    \centering
    \includegraphics[width=0.48\linewidth]{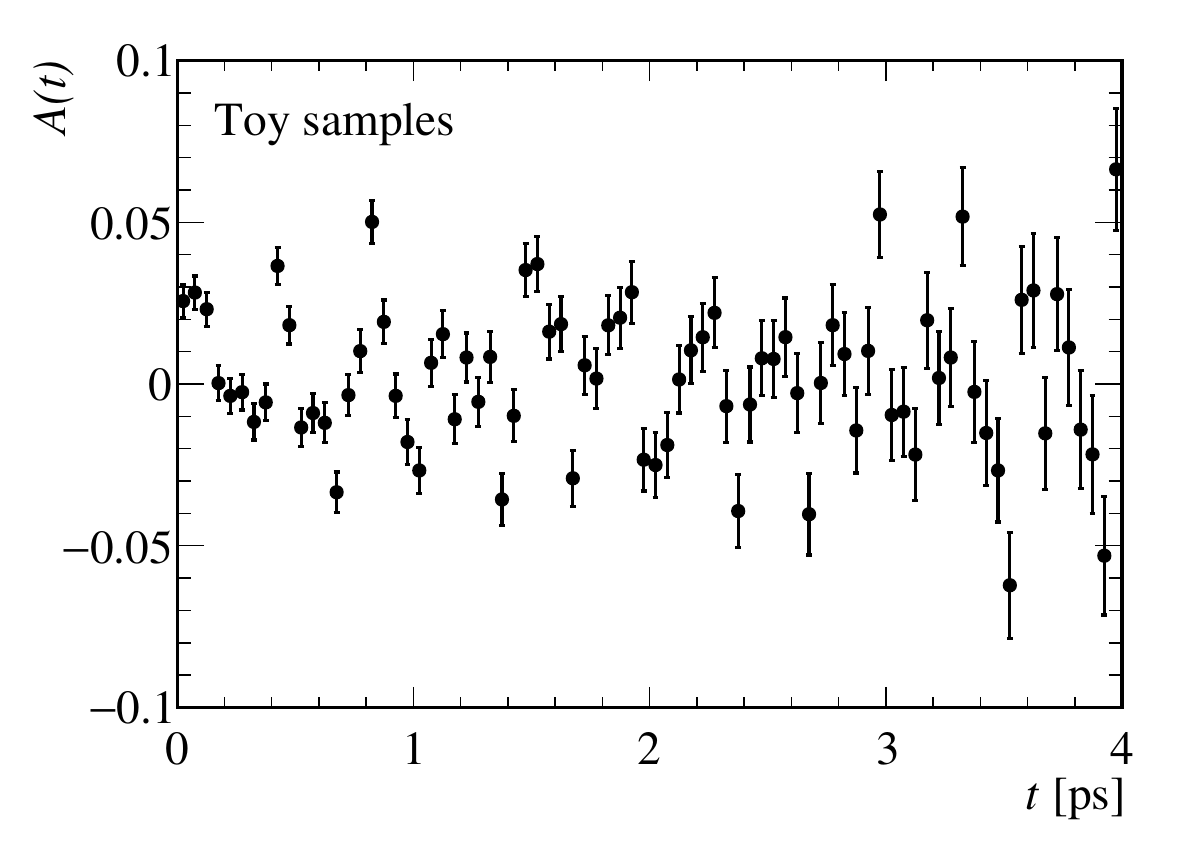} 
    \includegraphics[width=0.48\linewidth]{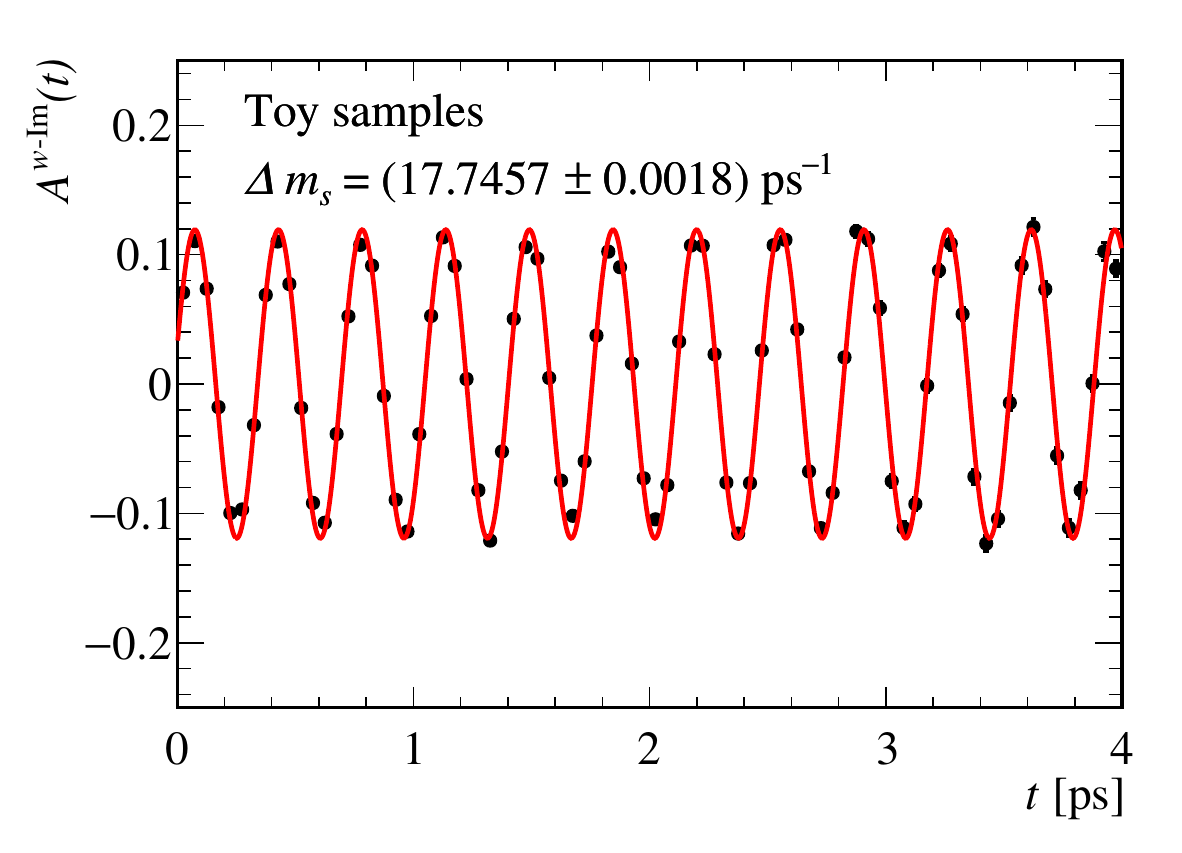}
    \includegraphics[width=0.48\linewidth]{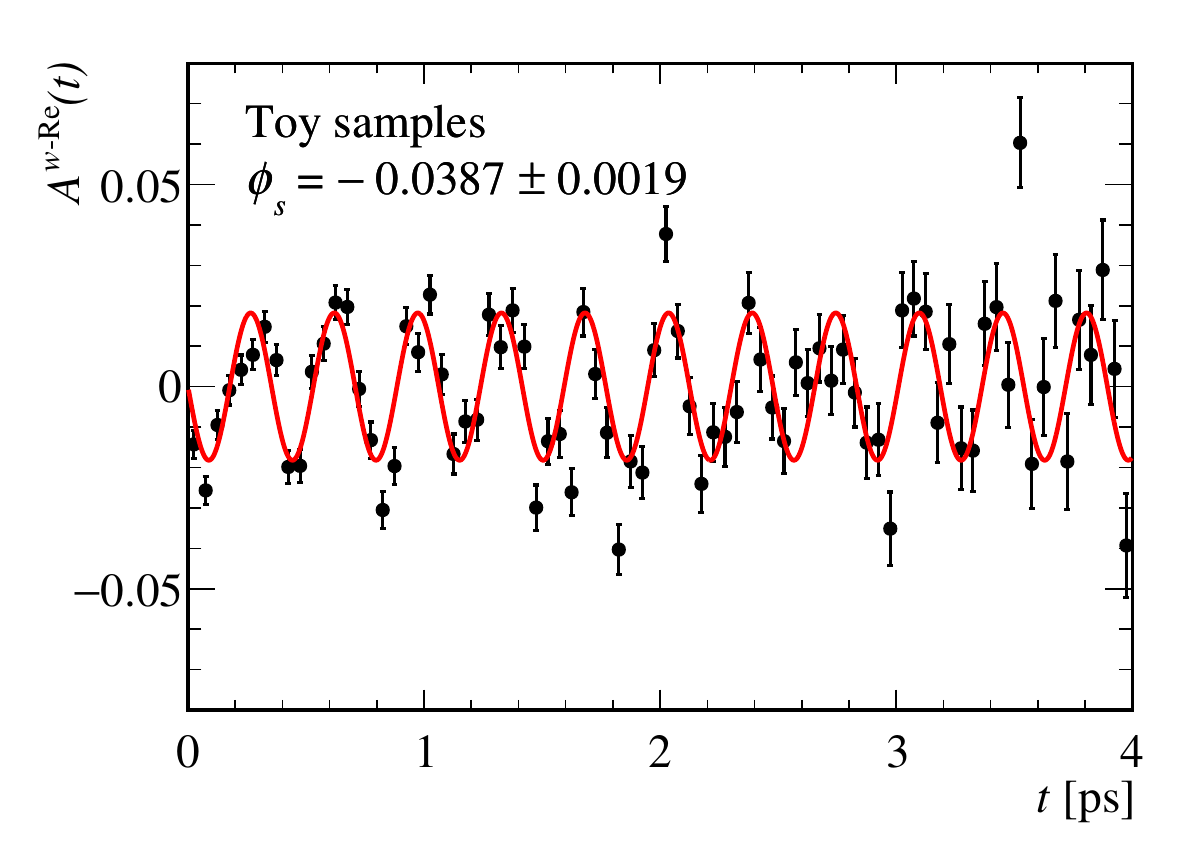}
    \includegraphics[width=0.48\linewidth]{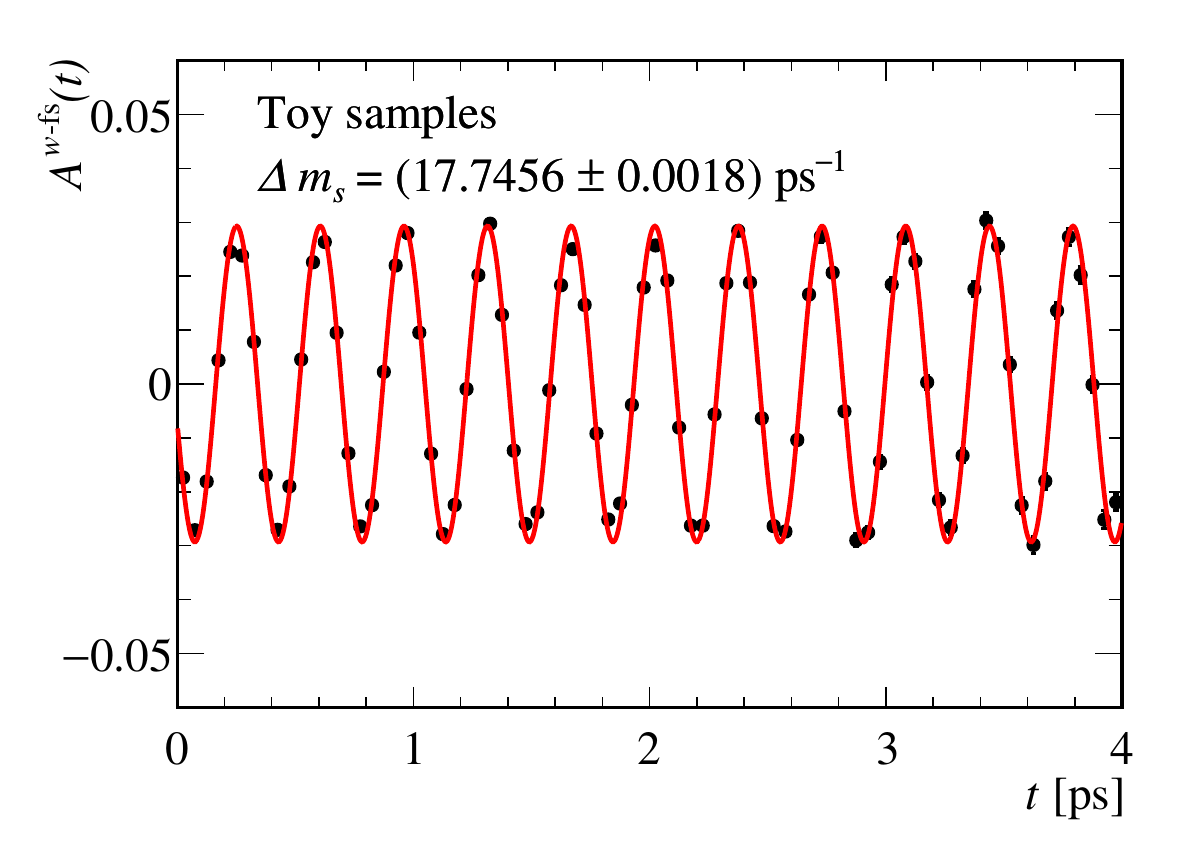}
    \caption{Asymmetry between the \Bsb- and \Bs-tagged decay-time distributions, for $\Bs \to \jpsi \phi$ decays after integrating over the decay angles: 
    (top left) without any weighting, (top right) weighted by $w^{\mathcal{I}m}$, (bottom left) weighted by $w^{\mathcal{R}e}$ and (bottom right) weighted by $w^{\rm fs}$.
    Curves described in the text are superimposed.
    Only the range $0 < t < 4\ps$ is shown to allow the oscillations to be seen.}
    \label{fig:Jpsipsi_toy}
\end{figure}

To demonstrate the impact of the different weighting functions, simulated $\Bs \to \jpsi \phi$ decays are generated.
The parameters used in the generation, listed in Table~\ref{tab:Jpsiphi-params}, are taken from the most recent LHCb publication~\cite{LHCb-PAPER-2023-016}, but without any \CP violation in decay and with the S-wave fraction set to zero for simplicity.
The sample size is approximately 100 times the corresponding yield for perfectly tagged events obtained from Ref.~\cite{LHCb-PAPER-2023-016}; a large sample size is used here to be ensure that \CP-violation effects suppressed by the small value of $\phi_s$ are visible.
Experimental effects are not included and initial-state flavour-tagging is performed with truth-level information.
The pseudodata are used to produce plots of the asymmetry between the \Bsb- and \Bs-tagged decay-time distributions, without any weighting and with each of the weighting functions $w^{\mathcal{I}m}$, $w^{\mathcal{R}e}$ and $w^{\rm fs}$ individually applied, as shown in Fig.~\ref{fig:Jpsipsi_toy}.
The asymmetries are fitted, in the range $0<t<10\ps$, with the functions 
\begin{eqnarray}    
    A^{w\mhyphen\mathcal{I}m}_{\CP}(t) & = & S^{w\mhyphen\mathcal{I}m} \sin(\Delta m_st) - C^{w\mhyphen\mathcal{I}m} \cos(\Delta m_st)\,, \nonumber \\
    && \text{where} \ S^{w\mhyphen\mathcal{I}m} = \cos(\phi_s) \times 0.1126 \ \text{and} \ C^{w\mhyphen\mathcal{I}m} = -0.0277\,, \label{eq:wIm-hadFactors} \\
    A^{w\mhyphen\mathcal{R}e}_{\CP}(t) & = & S^{w\mhyphen\mathcal{R}e} \sin(\Delta m_st)\,, \nonumber \\
    && \text{where} \ S^{w\mhyphen\mathcal{R}e} = -\sin(\phi_s) \times (-0.4606)\,, \label{eq:wRe-hadFactors} \\
    \text{and} \ A^{w\mhyphen{\rm fs}}_{\CP}(t) & = & S^{w\mhyphen{\rm fs}} \sin(\Delta m_st) - C^{w\mhyphen{\rm fs}} \cos(\Delta m_st)\,, \nonumber \\
    && \text{where} \ S^{w\mhyphen{\rm fs}} = \cos(\phi_s) \times (-0.0277) \ \text{and} \ C^{w\mhyphen{\rm fs}} = 0.0068\,, \label{eq:wfs-hadFactors} 
\end{eqnarray}
where the hadronic factors correspond to those in Eqs.~\eqref{eq:wS_fs-bilinear}--\eqref{eq:wC_Im-bilinear} and are evaluated with numerical integration neglecting the effect of non-zero $\Delta\Gamma_s$.
The hadronic factors are fixed to their known values in the fits to the asymmetries.
In the fits to $A^{w\mhyphen\mathcal{I}m}_{\CP}(t)$ and $A^{w\mhyphen{\rm fs}}_{\CP}(t)$ the value of $\cos(\phi_s)$ is also fixed, since there is little sensitivity to it, and the oscillation frequency $\Delta m_s$ is floated.
In the fit to $A^{w\mhyphen\mathcal{R}e}_{\CP}(t)$, the value of $\Delta m_s$ is fixed while the \CP violation parameter $\sin(\phi_s)$ is floated.  

Oscillations are visible in the unweighted data in Fig.~\ref{fig:Jpsipsi_toy}, due to the large sample of background free pseudodata and the non-zero value of the net-\CP of the final state (the net-\CP corresponds to $\left( |A_0|^2 + |A_\parallel|^2 - |A_\perp|^2 - |A_{\rm S}|^2 \right)/\left( |A_0|^2 + |A_\parallel|^2 + |A_\perp|^2 + |A_{\rm S}|^2 \right)$ which is equal to 0.5074 for the model used here).
Nonetheless, the oscillations in the $w^{\mathcal{R}e}$-weighted data are more clearly visible and demonstrate good sensitivity to $\sin\phi_s$.  
The oscillations in the $w^{\mathcal{I}m}$-weighted and $w^{\rm fs}$-weighted data are even more pronounced and show clearly the sensitivity to $\Delta m_s$.
Moreover, all fitted curves agree well with the weighted data and the fitted values are consistent with the inputs to the generation.

The illustration of the method presented here is done neglecting experimental effects, and therefore agreement of the pseudodata points with the fitted curve relies only on the correct evaluation of the hadronic factors.
In an analysis with real data, appropriate description of experimental effects such as background contributions will also be necessary to obtain agreement between fitted curves and the data.
In particular, imperfect initial-state flavour-tagging will reduce the magnitude of oscillations in all weighted asymmetries by an equal amount (assuming that the flavour-tagging performance is not correlated with position in the final-state phase space).  
The consistency of fitted curves with the data will therefore provide a powerful validation of the flavour-tagging algorithms that are crucial to obtain good sensitivity to decay-time-dependent \CP-violation effects.

\begin{table}[!tb]
    \centering
    \caption{Parameters used to generate $\Bs \to \jpsi \phi$ decays, based on the results from Ref.~\cite{LHCb-PAPER-2023-016}.}
    \label{tab:Jpsiphi-params}
    \begin{tabular}{c|c}
        \hline
        Parameter & Value \\
        \hline
        $\phi_s$ & $-0.039$ [rad] \\
        $|\lambda|$ & 1 \\ 
        $\Delta m_s$ & 17.743 [${\rm ps}^{-1}$]\\
        $\Gamma_s$ &   0.6523 [${\rm ps}^{-1}$] \\
        $\Delta\Gamma_s$ &  0.0845 [${\rm ps}^{-1}$] \\
        $|A_0|^2$ & 0.5179 \\
        $|A_\perp|^2$ & 0.2463 \\
        $|A_\parallel|^2$ & $1 - |A_0|^2 - |A_\perp|^2$\\
        $|A_{\rm S}|^2$ & 0 \\
        $\delta_\parallel-\delta_0$ &  3.146 [rad]\\
        $\delta_\perp-\delta_0$ & 2.903 [rad]\\
        $\delta_{\rm S}-\delta_0$ & 0\\
        \hline
    \end{tabular}
\end{table}

\subsubsection{Weighted untagged decay-time distributions}
\label{sec:illustration:Jpsiphi:untagged}

As described in Sec.~5 of Ref.~\cite{Gershon:2024xkk}, weighted untagged decay-time distributions can also provide useful visualisations.
This is most relevant for \Bs decays, since the interesting effects are related to the non-zero decay width difference, and would be particularly useful for experiments that do not have sufficient decay-time resolution to observe the fast \Bs oscillations. 
The untagged decay-time distribution is given by the sum of Eqs.~\eqref{eq:B0_to_fCP_wrt_time_ideal} and~\eqref{eq:B0bar_to_fCP_wrt_time_ideal}, 
\begin{equation}
\begin{array}{rcl}
    \Gamma_{\rm untagged}(t) & \propto &  \\
    && \hspace{-20mm} e^{-\Gamma t} 
        \int_{\rm PS}\left[ |\mathcal{A}_f|^2+|\overline{\mathcal{A}}_f|^2 \right] \cosh\left(\frac{\Delta\Gamma t}{2}\right) +2\mathcal{R}e\Big{(}e^{i\phi}\mathcal{A}_f^*\overline{\mathcal{A}}_f\Big{)} \sinh\left(\frac{\Delta\Gamma t}{2}\right)\,{\rm d}\Omega\,, \\
    && \hspace{-26mm} {}\propto e^{-t/\tau_{\Bs}}
        \int_{\rm PS}\sum_{k} N_k \left( a_k f_k(\Omega) \cosh\left(\frac{\Delta\Gamma_s t}{2}\right) + b_k f_k(\Omega) \sinh\left(\frac{\Delta\Gamma_s t}{2}\right)\right) \,{\rm d}\Omega\,, \\
    && \hspace{-26mm} {}\propto \frac{32\pi}{9} e^{-t/\tau_{\Bs}} 
        \left( \sum_{k \in (1,2,3,7)}\! N_k a_k \cosh\left(\frac{\Delta\Gamma_s t}{2}\right) - \cos\phi_s \!\sum_{k \in (1,2,3,7)}\! N_k \frac{b_k}{D} \sinh\left(\frac{\Delta\Gamma_s t}{2}\right)\right)\,. 
\end{array}\label{eq:Bs-untagged-Jpsiphi}
\end{equation}
The first line of Eq.~\eqref{eq:Bs-untagged-Jpsiphi} uses the notation of Secs.~\ref{sec:methodology} and~\ref{sec:illustration:Dpipi}, while the second and third use the notation for $\Bs \to \jpsi\phi$ decays introduced earlier in this subsection (recall the sign convention changes between $\Delta\Gamma$ and $\Delta\Gamma_s$ and between $\phi$ and $\phi_s$ discussed earlier).
The last line of Eq.~\eqref{eq:Bs-untagged-Jpsiphi} is after integration over the phase space and using $D = -\cos\phi_s$ to separate purely hadronic terms from those involving weak phase factors.
The integral here is the same as that of the denominator in Eqs.~\eqref{eq:S-wfs-fullDef} and~\eqref{eq:C-wfs-fullDef}, with $\sum_{k \in (1,2,3,7)} N_k a_k = \left( |A_0|^2 + |A_\parallel|^2 + |A_\perp|^2 + |A_{\rm S}|^2 \right)$ while $\sum_{k \in (1,2,3,7)} N_k \frac{b_k}{D} = \left( |A_0|^2 + |A_\parallel|^2 - |A_\perp|^2 - |A_{\rm S}|^2 \right)$.
For the model of $\Bs \to \jpsi\phi$ decays used in Sec.~\ref{sec:illustration:Jpsiphi:toy}, the former is equal to unity while the latter is 0.5074. 
As is well known, the distribution of Eq.~\eqref{eq:Bs-untagged-Jpsiphi} can be approximated by a single exponential function with effective lifetime deviating from the average $\Bs$ lifetime by an amount that depends on $\cos \phi_s$ and the net-\CP eigenvalue of the final state.  

Applying $w^{\mathcal{R}e}$ weights results in a weighted, untagged decay-time distribution given by
\begin{equation} \label{eq:Bs-wRe-untagged-Jpsiphi} 
\begin{array}{rcl}
    \Gamma^{w\mhyphen\mathcal{R}e}_{\rm untagged}(t) & \propto & e^{-\Gamma t}\left( 
        \int_{\rm PS}2\,\mathcal{R}e(\mathcal{A}_f^*\overline{\mathcal{A}}_f) \cosh\left(\frac{\Delta\Gamma t}{2}\right) 
        +\cos\phi \frac{\left[2\,\mathcal{R}e\left(\mathcal{A}_f^*\overline{\mathcal{A}}_f\right)\right]^2}{\left(|\mathcal{A}_f|^2+|\overline{\mathcal{A}}_f|^2\right)}\sinh\left(\frac{\Delta\Gamma t}{2}\right)\,{\rm d}\Omega
    \right)\,, \\ 
    & \propto & e^{-t/\tau_{\Bs}} 
        \int_{\rm PS}\sum_{k\in {\rm sym}} N_k \Bigg( \frac{d_k}{S} f_k(\Omega) \cosh\left(\frac{\Delta\Gamma_s t}{2}\right) + {}\\
    && \multicolumn{1}{r}{\frac{\sum_{k^\prime \in {\rm sym}}N_{k^\prime} b_k \frac{d_{k^\prime}}{S} f_k(\Omega)f_{k^\prime}(\Omega)}{\sum_{k^\prime \in {\rm sym}} N_{k^\prime} a_{k^\prime} f_{k^\prime}(\Omega)}\sinh\left(\frac{\Delta\Gamma_s t}{2}\right)\Bigg) \,{\rm d}\Omega\,,} \\
    & \propto & e^{-t/\tau_{\Bs}}
        \Bigg[ 
        \frac{32\pi}{9} \!\sum_{k \in (1,2,3,7)}\! N_k \frac{d_k}{S} \cosh\left(\frac{\Delta\Gamma_s t}{2}\right) - {} \\
    && \multicolumn{1}{r}{\cos \phi_s
    \int_{\rm PS} \frac{\sum_{k,k^\prime \in {\rm sym}}N_k N_{k^\prime} \frac{b_k}{D} \frac{d_{k^\prime}}{S} f_k(\Omega)f_{k^\prime}(\Omega)\,{\rm d}\Omega}{\sum_{k^\prime \in {\rm sym}} N_{k^\prime} a_{k^\prime} f_{k^\prime}(\Omega)}\sinh\left(\frac{\Delta\Gamma_s t}{2}\right)
    \Bigg]\,,}
\end{array}
\end{equation}
where again the second line uses the formalism of Sec.~\ref{sec:illustration:Jpsiphi:formalism} and the last equation is after trivial integrations and separating purely hadronic factors from weak phase information.
The hadronic factor $\sum_{k \in (1,2,3,7)} N_k \frac{d_k}{S} = -\sum_{k \in (1,2,3,7)} N_k \frac{b_k}{D} = -\left( |A_0|^2 + |A_\parallel|^2 - |A_\perp|^2 - |A_{\rm S}|^2 \right)$, so the first term would vanish in the limit of zero net-\CP.
Since $\frac{d_k}{S} = -\frac{b_k}{D}$ for $k \in {\rm sym}$, the hadronic factor of the second term in Eq.~\eqref{eq:Bs-wRe-untagged-Jpsiphi} is the same as that in Eq.~\eqref{eq:wS_Re-bilinear}, given in Eq.~\eqref{eq:wRe-hadFactors}, up to a factor of $\frac{32\pi}{9}\left( |A_0|^2 + |A_\parallel|^2 + |A_\perp|^2 + |A_{\rm S}|^2 \right)$, which for the model of $\Bs \to \jpsi\phi$ decays discussed in Sec.~\ref{sec:illustration:Jpsiphi:toy} is equal to $\frac{32\pi}{9} \approx 11.17$.

Applying the asymmetric weighting functions $w^{\mathcal{I}m}$ and $w^{\rm fs}$ has a larger effect on the untagged decay-time distribution, since the term proportional to $\cosh\left(\frac{\Delta\Gamma_s t}{2}\right)$ --- which, without weights, is fully symmetric with respect to charge conjugation of the final state --- vanishes after weighting and integration over the phase space.
This leaves distributions proportional to $e^{-t/\tau_{\Bs}}\sinh\left(\frac{\Delta\Gamma_s t}{2}\right)$, with magnitudes proportional to $\sin\phi_s$,
\begin{equation}\label{eq:Bs-wIm-untagged-Jpsiphi}
\begin{array}{rcl}
    \Gamma^{w\mhyphen\mathcal{I}m}_{\rm untagged}(t) 
    & \propto & e^{-\Gamma t}\left( 
        -\sin\phi\int_{\rm PS} \frac{\left[2\,\mathcal{I}m\left(\mathcal{A}_f^*\overline{\mathcal{A}}_f\right)\right]^2}{\left(|\mathcal{A}_f|^2+|\overline{\mathcal{A}}_f|^2\right)}\sinh\left(\frac{\Delta\Gamma t}{2}\right)\,{\rm d}\Omega
    \right)\,, \\ 
    & \propto & e^{-t/\tau_{\Bs}} 
        \int_{\rm PS}\sum_{k\in {\rm asym}} N_k \frac{\sum_{k^\prime \in {\rm asym}}N_{k^\prime} b_k \frac{d_{k^\prime}}{D} f_k(\Omega)f_{k^\prime}(\Omega)}{\sum_{k^\prime \in {\rm sym}} N_{k^\prime} a_{k^\prime} f_{k^\prime}(\Omega)}\sinh\left(\frac{\Delta\Gamma_s t}{2}\right) \,{\rm d}\Omega\,, \\
    & \propto & -\sin\phi_s 
        \int_{\rm PS} 
        \frac{\sum_{k,k^\prime \in {\rm asym}}N_K N_{k^\prime} \frac{b_k}{S} \frac{d_{k^\prime}}{D} f_k(\Omega)f_{k^\prime}(\Omega)\,{\rm d}\Omega}{\sum_{k^\prime \in {\rm sym}} N_{k^\prime} a_{k^\prime} f_{k^\prime}(\Omega)}\,e^{-t/\tau_{\Bs}}\sinh\left(\frac{\Delta\Gamma_s t}{2}\right)\,, 
\end{array}
\end{equation}
and
\begin{equation}\label{eq:Bs-wfs-untagged-Jpsiphi}
\begin{array}{rcl}        
    \Gamma^{w\mhyphen{\rm fs}}_{\rm untagged}(t) 
    & \propto & e^{-\Gamma t}\left( 
        -\sin\phi\int_{\rm PS} \frac{2\,\mathcal{I}m\left(\mathcal{A}_f^*\overline{\mathcal{A}}_f\right)\left(|\mathcal{A}_f|^2-|\overline{\mathcal{A}}_f|^2\right)}{\left(|\mathcal{A}_f|^2+|\overline{\mathcal{A}}_f|^2\right)}\sinh\left(\frac{\Delta\Gamma_s t}{2}\right)\,{\rm d}\Omega
    \right)\,, \\ 
    & \propto & e^{-t/\tau_{\Bs}} 
        \int_{\rm PS}\sum_{k\in {\rm asym}} N_k \frac{\sum_{k^\prime \in {\rm asym}}N_{k^\prime} b_k c_{k^\prime} f_k(\Omega)f_{k^\prime}(\Omega)}{\sum_{k^\prime \in {\rm sym}} N_{k^\prime} a_{k^\prime} f_{k^\prime}(\Omega)}\sinh\left(\frac{\Delta\Gamma_s t}{2}\right) \,{\rm d}\Omega\,, \\
    & \propto & -\sin\phi_s\,
        \int_{\rm PS} 
        \frac{\sum_{k,k^\prime \in {\rm asym}}N_k N_{k^\prime} \frac{b_k}{S} c_{k^\prime} f_k(\Omega)f_{k^\prime}(\Omega)\,{\rm d}\Omega}{\sum_{k^\prime \in {\rm sym}} N_{k^\prime} a_{k^\prime} f_{k^\prime}(\Omega)}\,e^{-t/\tau_{\Bs}}\sinh\left(\frac{\Delta\Gamma_s t}{2}\right)\,.
\end{array}
\end{equation}
Since $\frac{d_k}{D} = \frac{b_k}{S}$ for $k \in {\rm asym}$, the hadronic factors in 
Eqs.~\eqref{eq:Bs-wIm-untagged-Jpsiphi} and~\eqref{eq:Bs-wfs-untagged-Jpsiphi} are the same as those in Eqs.~\eqref{eq:wS_Im-bilinear} and~\eqref{eq:wC_Im-bilinear}, respectively, up to factors of $\frac{32\pi}{9}\left( |A_0|^2 + |A_\parallel|^2 + |A_\perp|^2 + |A_{\rm S}|^2 \right)$.
The numerical values evaluated for the model of $\Bs \to \jpsi\phi$ decays discussed above are given in Eq.~\eqref{eq:wIm-hadFactors}; in this model the second factor is about a factor of four smaller than the first.

\begin{figure}[!tb]
    \centering
    \includegraphics[width=0.48\linewidth]{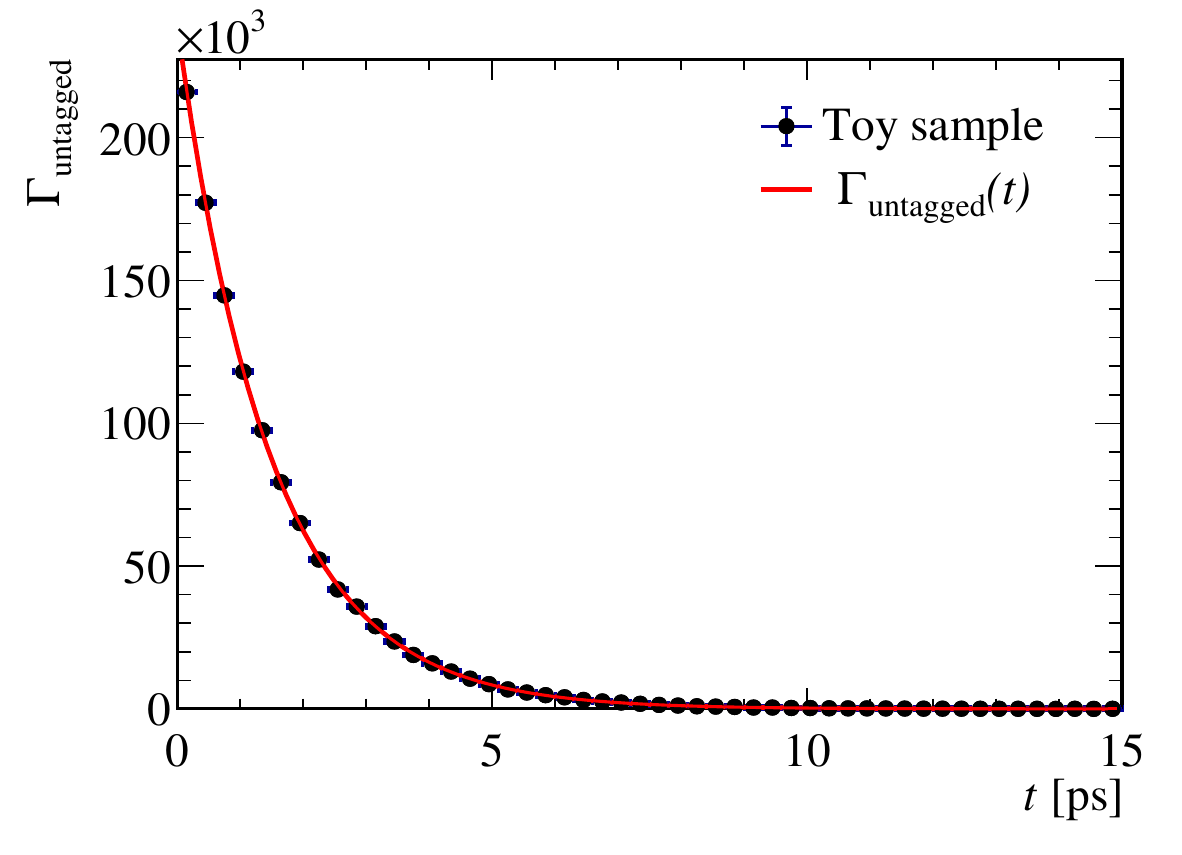} 
    \includegraphics[width=0.48\linewidth]{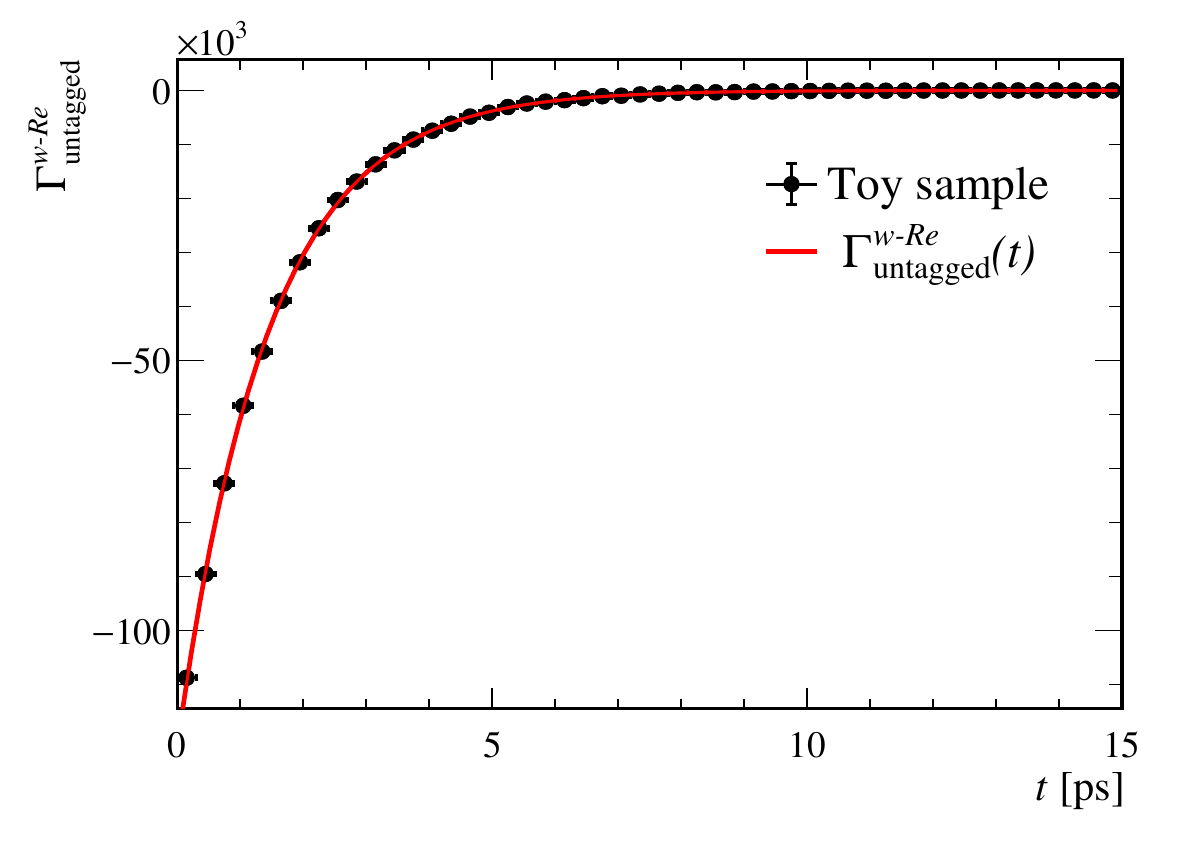}
    \includegraphics[width=0.48\linewidth]{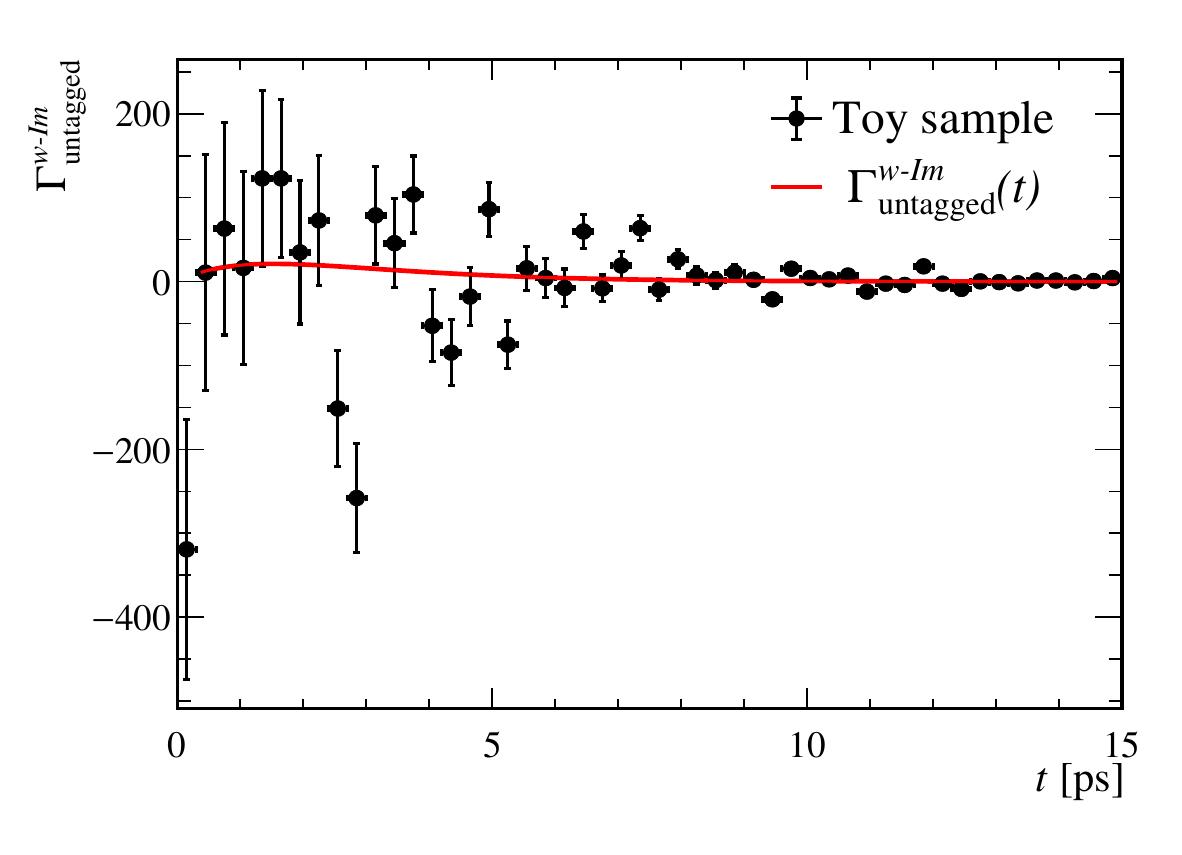}    
    \includegraphics[width=0.48\linewidth]{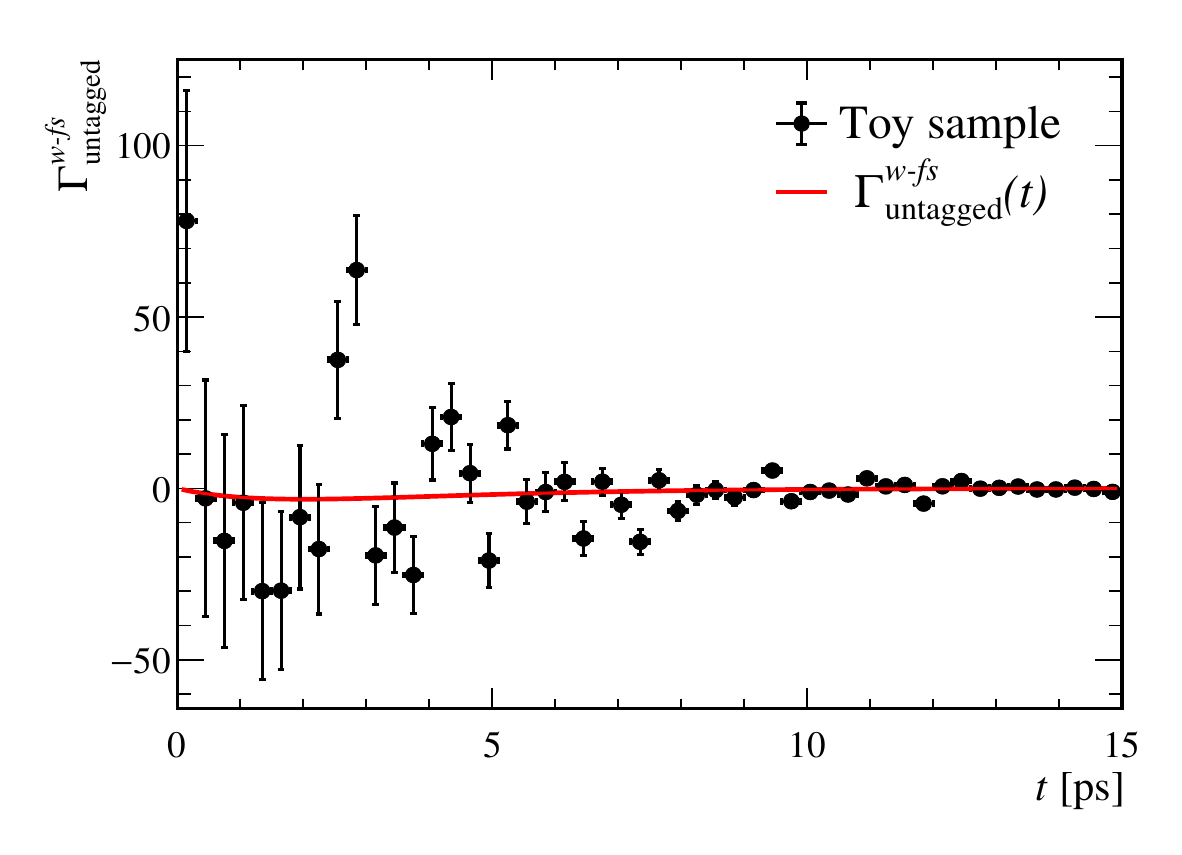}
    \caption{Distributions of untagged decay-rate distributions 
    (top left) without weighting, 
    and (top right) with $w^{\mathcal{R}e}$, (bottom left) $w^{\mathcal{I}m}$ weights applied, and (bottom right) $w^{\rm fs}$ weights applied, for the pseudoexperiments for $\Bs \to \jpsi \phi$ decays.
    Curves corresponding to the expected distributions as given in Eqs.~\eqref{eq:Bs-untagged-Jpsiphi}--\eqref{eq:Bs-wIm-untagged-Jpsiphi} are also shown.
    }
    \label{fig:Jpsipsi_toy_time}
\end{figure}

The untagged decay-time distributions, both without and with weights applied, for the same sample of simulated $\Bs \to \jpsi \phi$ decays used in Sec.~\ref{sec:illustration:Jpsiphi:toy} are shown in Fig.~\ref{fig:Jpsipsi_toy_time}.
The unweighted distribution and that obtained after applying $w^{\mathcal{R}e}$ weights are, as expected, close to exponential decay distributions, albeit with a negative normalisation in the latter case.
It was noted in Ref.~\cite{Gershon:2024xkk} that the weighted untagged decay-time distribution corresponding to Eq.~\eqref{eq:Bs-wIm-untagged-Jpsiphi}, which provides visualisation of the emergence, as a function of decay time, of an asymmetry across the Dalitz-plot, provides a novel way to observe \CP\ violation in the interference between mixing and decay.
This can also be seen for the $w^{\mathcal{I}m}$ weighted data in Fig.~\ref{fig:Jpsipsi_toy_time}, although the small value of $\phi_s$ makes the effect insignificant even for the large sample used in this study.  
As shown here for the first time, the untagged decay-time distribution obtained applying $w^{\rm fs}$ weights also provides a similar possibility, although for the model of $\Bs \to \jpsi\phi$ decays based on results from the LHCb collaboration~\cite{LHCb-PAPER-2023-016} it is somewhat less sensitive.

\section{Summary}
\label{sec:summary}

Visualisation of decay-time-dependent asymmetries in neutral \B\ meson decays to multibody final states is valuable as it enables simple demonstrations of the agreement of fit results with the multidimensional data.  
The methods proposed in Ref.~\cite{Gershon:2024xkk} and here to achieve this allow both \CP-violating and \CP-conserving asymmetries to be visualised.  
While the former are likely to be of most interest in order to test the Standard Model of particle physics, the latter are also highly useful to test the agreement of the fit model with data, in particular to validate the performance of algorithms used to tag the flavour of the initially produced neutral \B meson.  
The formalism presented in Ref.~\cite{Gershon:2024xkk} to achieve this in terms of total amplitudes for \B and \Bbar decays is generic, but the relevant formulae can also be presented in terms of bilinear combinations of transversity amplitudes as commonly used for \B meson decays to two vector mesons. 
The demonstration of the method in this paper is expected to be useful to facilitate implementation of the method in future experimental studies of $\Bs \to \jpsi \phi$ decays and other similar processes.

\section*{Acknowledgements}

The authors wish to thank their colleagues on the LHCb experiment for the fruitful and enjoyable collaboration. 
In particular, they would like to thank Veronika Chobanova and Stefania Vecchi for helpful comments.
TG, TL and MW are supported by the Science and Technology Facilities Council (UK).
AM acknowledges support from the European Research Council Starting grant ALPACA 101040710.
PL and WQ are supported by Fundamental Research Funds for the Central Universities.
WQ is also supported by the National Science Foundation of China under Grant No.\ 11975015, and the National Key R\&D Program of China under Grant No.\ 2022YFA1601901.

\addcontentsline{toc}{section}{References}
\setboolean{inbibliography}{true}
\bibliographystyle{LHCb}
\bibliography{references,standard,main,LHCb-PAPER,LHCb-CONF,LHCb-DP,LHCb-TDR}

\end{document}